\documentclass[letterpaper,12pt]{article}

\usepackage{graphics} 
\usepackage{times} 
\usepackage{amsmath} 
\usepackage{amssymb}  
\usepackage{amsthm}

\usepackage{float}
\usepackage{graphicx}
\usepackage{xcolor}

\title{\LARGE \bf
Baseband Equivalent Models and Digital Predistortion for Mitigating 
Dynamic Continuous-Time Perturbations in Phase-Amplitude 
Modulation-Demodulation Schemes 
\thanks{This work was supported by DARPA Award No. 
W911NF-10-1-0088. Shorter version of this paper was published in 
the proceedings of the 55th IEEE Conference on Decision and 
Control \cite{Tanovic2016}}
}

\author{Omer Tanovic \thanks{Omer Tanovic and Alexandre Megretski are with the 
Laboratory for Information and Decision Systems (LIDS), Department of Electrical Engineering 
and Computer Science (EECS), Massachusetts Institute of Technology (MIT), Cambridge, MA 02139, 
USA {\tt\small \{otanovic,ameg\}@mit.edu}}, Alexandre Megretski\footnotemark[2], 
Yan Li \thanks{Yan Li was with the Research Laboratory of Electronics, 
EECS, MIT. She is now with NanoSemi Inc., Waltham, MA 02451, USA 
{\tt\small yan.li@nanosemitech.com}}, \\
Vladimir M. Stojanovic \thanks{Vladimir M. Stojanovic is with the Department of Electrical 
Engineering and Computer Sciences, University of California Berkeley, Berkeley, 
CA 94720, USA {\tt\small vlada@berkeley.edu}}, 
and Mitra Osqui \thanks{Mitra Osqui was with LIDS, EECS, MIT. She is now with 
Analog Devices $|$ Lyric Labs, Cambridge, MA 02142, USA {\tt\small mitra.osqui@analog.com}}
}

\begin{document}

\newcommand{\MRK}[1]{{\bf[#1]}}
\newcommand{\rf}[1]{(\ref{#1})}

\maketitle
\thispagestyle{empty}
\pagestyle{empty}

\begin{abstract}
We consider baseband equivalent representation of transmission circuits, in the form 
of a nonlinear dynamical system $\mathbf S$ in discrete time (DT) defined by a series 
interconnection of a phase-amplitude modulator, a nonlinear dynamical system $\mathbf F$
in continuous time (CT), and an ideal demodulator.
We show that when $\mathbf F$ is a CT Volterra series model, the resulting $\mathbf S$ 
is a series interconnection of a DT Volterra series model of same degree and memory depth, 
and an LTI system with special properties. The result suggests a new, non-obvious, 
analytically motivated  structure of digital pre-compensation of analog nonlinear 
distortions such as those caused by power amplifiers in digital communication systems. 
The baseband model and the corresponding digital compensation structure readily 
extend to OFDM modulation.  MATLAB simulation is used to verify proposed baseband 
equivalent model and demonstrate effectiveness of the new compensation scheme, as 
compared to the standard Volterra series approach. 

\end{abstract}
\vskip5mm

\section{Notation and Terminology}
\noindent $j$ is a fixed square root of $-1$.
$\mathbb C$, $\mathbb R$, $\mathbb Z$ $\mathbb N$ are the standard sets
of complex, real, integer, and positive integer numbers.
$X^d$, for a set $X$, is the set of all $d$-typles
$(x_1,\dots,x_d)$ with $x_i\in X$. For a set $S$,
$|S|$ denotes the number of elements in $S$ ($|S|=\infty$ when $S$ is
not finite).
\vskip1mm
\noindent In this paper, (scalar) {\sl CT signals} are 
uniformly bounded square integrable functions $\mathbb R\to\mathbb R$.
The set of all CT signals is denoted by $\mathcal L$.
$n$-dimensional {\sl DT signals} are the elements of $\ell_n$
(or simply $\ell$ for $n=1$), the set of all square summable 
functions $\mathbb Z\to\mathbb C^n$. 
For $w\in\ell$, $w[n]$ denotes
the value of $w$ at $n\in\mathbb Z$. 
In contrast, $x(t)$ refers to the value of $x\in\mathcal L$ at $t\in\mathbb R$.
The Fourier transform $\mathcal F$ applies to both CT and DT  signals.
For $x\in\mathcal L$, its Fourier transform  $X=\mathcal Fx$
is a square integrable function 
$X:~\mathbb R\to\mathbb C$. For
$x\in\ell_n$, the Fourier transform $X=\mathcal Fx$ is a $2\pi$-periodic function
$X:~\mathbb R\to\mathbb C$,  square integrable on its period.

\vskip1mm
\noindent {\sl Systems} are viewed as functions $\mathcal L\to\mathcal L$,
$\mathcal L\to\ell$, $\ell\to\mathcal L$, or $\ell_k\to\ell_m$. $\mathbf Gf$
denotes the response 
of system $\mathbf G$ to signal $f$ 
(even when $\mathbf G$ is not linear), and the {\sl series
composition} $\mathbf K=\mathbf{QG}$ of systems
 $\mathbf Q$ and $\mathbf G$ is the system mapping $f$ to
$\mathbf{Q}(\mathbf{G}f)$.
A system $\mathbf G:~L\to L$ 
(or $\mathbf G:~\ell\to\ell$) is said to be {\sl linear and time invariant} (LTI) with
{\sl frequency response} $H:~\mathbb R\to\mathbb C$ when
$\mathcal F\mathbf Gx = H\cdot\mathcal F\mathbf x$ for all $x\in\mathcal L$
(respectively $x\in\ell$).

\section{Introduction}
In modern communications systems, with demand for high-throughput data 
transmission, requirements on the system linearity become more strict. This 
is in large part due to a combination of ever increasing signalling rates with 
use of more complex modulation/demodulation schemes for enhanced 
spectral efficiency. This in turn forces RF transmitter power amplifiers (PA) to 
operate over a large portion of their transfer curves, generating out of band 
spectral content which degrades spectral efficiency. A common way to make 
the PA (and correspondingly the whole signal chain) behave linearly is to 
back-off PA's input level, which results in reduced power efficiency \cite{Ken2000}. 
This motivates the search for a method which would help increase both linearity 
and power efficiency. Digital compensation offers an attractive approach to 
designing electronic devices with superior characteristics, and it is not a surprise 
that it has been used in PA linearization as well. 
Nonlinear distortion in an analog system can be compensated with a pre-distorter 
or a post-compensator system. This pre-distorter inverts nonlinear behavior of 
the analog part, and is usually implemented as a digital system. Techniques 
which employ such systems are called digital predistortion (DPD) techniques, 
and they can produce highly linear transmitter circuits 
\cite{Ken2000}-\cite{Vuolevi2003}. 

DPD structure usually depends on behavioral PA models and their baseband 
equivalent counterparts. First attempts to mitigate PA's nonlinear effects by 
employing DPD involved using simple memoryless models in order to describe 
PA's behavior \cite{Saleh1983}. As the signal bandwidth has increased over 
time, it has been recognized that short and long memory effects play significant 
role in PA's behavior \cite{Bosch1989}, and should be incorporated into the 
model. Since then several memory baseband models and corresponding 
predistorters have been proposed to compensate memory effects: memory 
polynomials \cite{Kim2001,Ding2004}, Hammerstein and Wiener models 
\cite{Mathews2000}, pruned Volterra series \cite{Zhu2004}, generalized 
memory polynomials \cite{Morgan2006}, dynamic deviation reduction-based 
Volterra models \cite{Zhu2006, Zhu2008}, as well as the most recent neural 
networks based behavioral models \cite{Rawat2010}, and generalized rational 
functions based models \cite{Rawat2014}. These papers emphasize capturing 
the whole range of the output signal's spectrum, which is proportional to the 
order of nonlinearity of the RF PA, and is in practice taken to be about five 
times the input bandwith {\textcolor{red}{[???]}}. In wideband communication systems this would make 
the linearization bandwidth very large, and hence would put a significant 
burden on the system design (e.g., would require very high-speed data 
converters). Since these restrictions limit applicability of conventional models 
in the forthcoming wideband systems \textcolor{red}{(e.g. LTE-advanced)}, it is beneficial 
to investigate model dynamics when the PA's output is also limited in bandwidth. 
In that case DPD would ideally mitigate distortion in that frequency band, and 
possible adjacent channel radiation could be taken care of by applying 
bandpass filter to the PA's output. Such band-limited baseband model and its 
corresponding DPD were investigated in \cite{Yu2012}, and promising 
experimental results were shown. 
Theoretical analysis shown in \cite{Yu2012} follows the same modeling approach 
as the conventional baseband models (dynamic deviation reduction-based Volterra 
series modeling). Due to the bandpass filtering operation applied on the PA 
output, long (possibly 
infinite) memory dynamic behavior is now present, which makes these band-limited 
models fundamentally different from the conventional baseband models. Hence standard 
modeling methods, such as memory polynomials or dynamic deviation reduction-based 
Volterra series modeling, might be too general to pinpoint this new structure, 
and also not well suited for practical 
implementations (long memory requirements in nonlinear models would require 
exponentially large number of coefficients to be implemented 
\textcolor{red}{in e.g. look-up table models).} 

In this paper, we develop an explicit expression of the equivalent baseband model, 
when the passband nonlinearity can be described by a Volterra series model with 
fixed degree and memory depth. We show that this baseband model can be written 
as a series interconnection of a fixed degree and short memory discrete Volterra model, 
and a long memory discrete LTI system which can be viewed as a bank of 
{\sl reconstruction filters}. In other words, we show that the underlying baseband 
equivalent structure alows for untangling of passband nonlinearity, of relatively short 
memory, and long memory requirements imposed by bandpass filtering and 
modulation/demodulation operation. In exact analytical representation, the above 
reconstruction filters exibit 
discontinuities at frequency values $\pm\pi$, making their unit step responses infinitely 
long. Nevertheless, the reconstruction filters are shown to be smooth inside the interval 
$(-\pi,\pi)$, and thus aproximable by low order FIR filters. Both relatively low 
memory/degree requirements of the nonlinear (Volterra) subsystem and good 
approximability by FIR filters of the linear subsystem, allow for potentially efficient 
hardware implementation of the corrresponding baseband model. 
Suggested by the derived model, we propose a non-obvious, 
analytically motivated structure of digital precompensation of RF PA nonlinearities. 

This paper is organized as follows. In Section III we further discuss motivation for 
considering problem of baseband equivalent modeling and digital predistortion, and 
give mathematical description 
of the system under consideration. Main result is stated and proven in Section IV, 
i.e. in this section we give an explicit expression of the equivalent baseband model. 
In Section V we provide some further discussion on advantages of the proposed method, 
and its extension to OFDM modulation. 
DPD design and its performance are demonstrated by  MATLAB simulation results
presented in Section VI. 

\section{Motivation and Problem Setup}

In this paper, a digital compensator is viewed as a system 
$\mathbf C:~\ell\to\ell$. More specifically,
a {\sl pre-}compensator $\mathbf C:~\ell\to\ell$
designed for a device modeled by a system 
$\mathbf S:~\ell\to\mathcal L$ (or
$\mathbf S:~\ell\to\ell$) aims to
make the composition $\mathbf{SC}$, as shown on the block diagram below,
\vskip3mm
\setlength{\unitlength}{0.09cm}
\begin{center}\begin{picture}(64,8)(14,50)
\put(26,50){\framebox(12,8){$ \mathbf C$}}
\put(52,50){\framebox(12,8){$ \mathbf S$}}
\put(14,54){\vector(1,0){12}}
\put(38,54){\vector(1,0){14}}
\put(64,54){\vector(1,0){12}}
\put(14,56){$u$}
\put(44,56){$w$}
\put(76,56){$v$}
\end{picture}\end{center}
\noindent conform to a set of desired specifications.
(In the simplest scenario, the objective is to make $\mathbf{SC}$
as close to the identity map as possible,
in order to cancel the distortions introduced by 
$\mathbf S$.)

A common element in digital compensator design algorithms is
selection of {\sl compensator structure}, which usually means
specifying a finite sequence $\tilde{\mathbf C}=(\mathbf C_1,\dots\mathbf C_N)$
of systems $\mathbf C_k:~\ell\to\ell$, and restricting the
actual compensator $\mathbf C$ to have the form
\[ \mathbf C=\sum_{k=1}^N a_k\mathbf C_k,\qquad a_k\in\mathbb C,\]
i.e., to be a linear combination of the elements of $\tilde{\mathbf C}$.
Once the {\sl basis} sequence $\tilde{\mathbf C}$ is fixed, the design usually reduces to
a straightforward {\sl least squares optimization} of the coefficients $a_k\in\mathbb C$.

A popular choice is for the systems $\mathbf C_k$ to be some {\sl Volterra monomials},
i.e. to map their input $u=u[n]$ to the outputs $w_k=w_k[n]$ according to the polynomial 
formulae
\[  w_k[n] = \prod_{j=1}^{d_r(k)}\text{Re}~u[n-n^r_{k,j}]
\prod_{j=1}^{d_i(k)}\text{Im}~u[n-n^i_{k,j}],
\]
where the integers $d_r(k)$ and $d_i(k)$ (respectively, $n^r_{k,j}$ and $n^i_{k,j}$) 
will be referred to as {\sl degrees} (respectively, {\sl delays}). 
In this case, every linear combination $\mathbf C$ of $\mathbf C_k$ is a
{\sl DT Volterra series}  \cite{Schetz2006}, i.e., a  DT system
mapping signal inputs $u\in\ell$ to outputs $w\in\ell$ 
according to the polynomial expression
\[ w[n]=\sum_{k=1}^{N}a_k\prod_{j=1}^{d_r(k)}\text{Re}~u[n-n^r_{k,j}]
\prod_{j=1}^{d_i(k)}\text{Im}~u[n-n^i_{k,j}].\]

\noindent 
Selecting a proper {\sl compensator structure} is a major challenge
in compensator design: a basis
 which is too simple will not be capable of cancelling the distortions
well, while a form that is too complex will consume excessive power
and space. Having an insight into the compensator
basis selection can be very valuable. For  an example (cooked up outrageously
to make the point), consider the case when the ideal 
compensator $\mathbf C:~u\mapsto w$ is
given by
\[  w[n] = \rho u[n]+\delta\left(\sum_{j=-50}^{50}u[n-j]\right)^5\]
for some (unknown) coefficients $\rho$ and $\delta$.
One can treat $\mathbf C$ as a generic Volterra series expansion
with fifth order monomials with delays between $-50$ and $50$, and the first order monomial
with delay 0, which leads to a basis sequence $\tilde{\mathbf C}$ with
$1+{105\choose 5}=96560647$ elements (and the same number of multiplications involved
in implementing the compensator). Alternatively, one may realize that
the two-element structure $\tilde{\mathbf C}=\{\mathbf C_1,\mathbf C_2\}$, with
$w_k=\mathbf C_k u$ defined by
\[  w_1[n]=u[n],\qquad w_2[n]=\left(\sum_{j=-50}^{50}u[n-j]\right)^5\]
is good enough.
 
In this paper we establish that a certain special 
structure is good enough to compensate for
imperfect modulation. We consider modulation systems 
represented by the block diagram

\setlength{\unitlength}{0.08cm}
\begin{center}\begin{picture}(84,15)(8,64)
\put(24,64){\framebox(16,12){$ \mathbf M$}}
\put(58,64){\framebox(16,12){$ \mathbf F$}}
\put(8,70){\vector(1,0){16}}
\put(40,70){\vector(1,0){18}}
\put(74,70){\vector(1,0){18}}
\put(8,72){$u[n]$}
\put(48,72){$x(t)$}
\put(92,72){$y(t)$}
\end{picture}\end{center}
where $\mathbf M:~\ell\to\mathcal L$ is the 
{\sl ideal} modulator, and $\mathbf F:~\mathcal L\to\mathcal L$ 
is a CT dynamical system used to represent linear and 
nonlinear distortion in the modulator and power amplifier circuits.
We consider the ideal modulator of the form
 $\mathbf M=\mathbf X\mathbf Z$, where
$\mathbf Z:~\ell\to\mathcal L$ is the {\sl zero order hold}
map $u[\cdot]\mapsto x_0(\cdot)$:
\begin{equation} \label{mod_eq}
\quad x_0(t)=\sum_n p(t-nT)u[n],\quad
p(t)=\begin{cases}1,& t\in[0,T),\\ 0,& t\not\in[0,T)\end{cases}
\end{equation}
with fixed sampling interval length $T>0$
and $\mathbf X:~\mathcal L\mapsto\mathcal L$ is the
{\sl mixer} map
\[ x_0(\cdot)\mapsto x(\cdot):\quad x(t)=2\text{Re}[\exp(j\omega_ct)x_0(t)] \]
with modulation-to-sampling frequency ratio $M\in\mathbb N$, i.e., with
$\omega_c=2\pi M/T$.
 We are particularly interested in the case when $\mathbf F$ 
is described by the {\sl CT Volterra series model}
\begin{equation}\label{ctvs}
  y(t)=b_0+\sum_{k=1}^{N_b}b_k\prod_{i=1}^{\beta_k}x(t-t_{k,i}),
\end{equation}
where $N_b\in\mathbb N$, $b_k\in\mathbb R$, $\beta_k\in\mathbb N$, $t_{k,i}\ge0$
are parameters. (In a similar fashion,  it is possible to consider input-output relations
in which the finite sum in (\ref{ctvs}) is replaced by an integral, or an infinite sum).
One expects that the memory of $\mathbf F$ is not long, compared to $T$,
i.e., that $\max t_{k,i}/T$ is not much larger than 1.

As a rule, the spectrum of the DT input $u\in\ell$ of the modulator is carefully 
shaped at a pre-processing stage
to guarantee desired characteristics of the modulated signal 
$x=\mathbf Mu$. However,
when the distortion $\mathbf F$ is not linear, the spectrum of the
$y=\mathbf Fx$ could be damaged substantially, 
leading to violations of  EVM and spectral mask
specifications \cite{Zhu2008}. 

Consider the possibility of repairing the spectrum of $y$ by pre-distorting the digital input
$u\in\ell$ by a compensator 
$\mathbf C:~\ell\to\ell$, 
as shown
on the block diagram below:
\setlength{\unitlength}{0.11cm}
\begin{center}\begin{picture}(72,10)(4,60)
\put(14,60){\framebox(12,8){$ \mathbf C$}}
\put(34,60){\framebox(12,8){$ \mathbf M$}}
\put(54,60){\framebox(12,8){$\mathbf F$}}
\put(4,64){\vector(1,0){10}}
\put(26,64){\vector(1,0){8}}
\put(46,64){\vector(1,0){8}}
\put(66,64){\vector(1,0){10}}
\put(47,66){$x(t)$}
\put(72,66){$y(t)$}
\put(4,66){$u[n]$}
\put(26.5,66){$w[n]$}
\end{picture}\end{center}

\noindent
The desired effect of inserting $\mathbf C$ 
is cancellation of the distortion caused by 
$\mathbf F$.
Naturally, since $\mathbf C$ acts in the baseband (i.e., in discrete time), 
there is no chance that
$\mathbf C$ will achieve a complete correction, i.e., that the series composition 
$\mathbf F\mathbf M\mathbf C$ of $\mathbf F$, 
$\mathbf M$, and $\mathbf C$ will
be identical to $\mathbf M$. 
However, in principle, it is sometimes possible to make the frequency
contents of $\mathbf Mu$ and $\mathbf F\mathbf M\mathbf Cu$ to be identical 
within the CT frequency band $(\omega_c-\omega_b,\omega_c+\omega_b)$, where
$\omega_b=\pi/T$ is the Nyquist frequency
\cite{Frank1996,Tsimbinos1998}.
To this end, let $\mathbf H:~\mathcal L\to\mathcal L$ 
denote the ideal band-pass filter with frequency response
\begin{equation} 
\label{bandpass}
 H(\omega)=\begin{cases}
 1,& |~\omega_c-|\omega|~|<\omega_b,\\ 
 0,& \text{otherwise}.
\end{cases}
\end{equation}
Let $\mathbf D:~\mathcal L\to\ell$ 
be the {\sl ideal de-modulator} relying on the band selected by 
$\mathbf H$, i.e. the linear system for which the
series composition $\mathbf D\mathbf H\mathbf M$ is the identity
function. 
Let $\mathbf S=\mathbf D\mathbf H\mathbf F\mathbf M$ be the series composition 
of $\mathbf D$, $\mathbf H$, $\mathbf F$, and $\mathbf M$, i.e. the DT system
with input $w=w[n]$ and output $v=v[n]$ shown on the block diagram below:

\begin{figure}[H]
\setlength{\unitlength}{0.1cm}
\begin{center}\begin{picture}(96,11)(2,58)
\put(12,58){\framebox(12,8){$\mathbf M$}}
\put(34,58){\framebox(12,8){$\mathbf  F$}}
\put(56,58){\framebox(12,8){$\mathbf  H$}}
\put(78,58){\framebox(10,8){$\mathbf  D$}}
\put(2,62){\vector(1,0){10}}
\put(24,62){\vector(1,0){10}}
\put(46,62){\vector(1,0){10}}
\put(68,62){\vector(1,0){10}}
\put(88,62){\vector(1,0){10}}
\put(2,64){$w[n]$}
\put(25,64){$x(t)$}
\put(47,64){$y(t)$}
\put(69,64){$z(t)$}
\put(94,64){$v[n]$}
\end{picture}\end{center}
\caption{Block diagram of $\bf S=\mathbf D\mathbf H\mathbf F\mathbf M$}
\label{fig:whole_chain}
\end{figure}
\noindent
By construction, the ideal compensator $\mathbf C$ should be the inverse
$\mathbf C=\mathbf S^{-1}$ of $\mathbf S$, as long as the inverse does exist.

A key question answered in this paper is "what to expect from system $\mathbf S$?"
If one assumes that the continuous-time distortion subsystem $\mathbf F$ is
simple enough, what does this say about $\mathbf S$?

This paper provides an explicit expression for $\mathbf S$ in the case when
$\mathbf F$ is given in the CT Volterra series form \rf{ctvs} with
{\sl degree} $d=\max\beta_k$ and {\sl depth} $t_{max}=\max t_{k,i}$. The 
result reveals that, even though $\mathbf S$ tends to have infinitely long memory
(due to the ideal band-pass filter $\mathbf H$ being involved in the construction 
of $\mathbf S$), it can be represented as a series composition 
$\mathbf S=\mathbf L\mathbf V$, where  
$\mathbf V:~\ell\to\ell_N$ maps scalar
complex input $w\in\ell$ to real vector output 
$g\in\ell_N$ in such a way that
the $k$-th scalar component $g_k[n]$ of $g[n]\in\mathbb R^N$
is given by
\[  g_k[n]=\prod_{i=0}^{m}(\text{Re}~w[n-i])^{\alpha_i}~
\prod_{i=0}^{m}(\text{Im}~w[n-i])^{\beta_i}, \quad\quad
\alpha_i,\beta_i\in\mathbb Z_+,\qquad
 \sum_{i=0}^m\alpha_i+\sum_{i=0}^m\beta_i\le d,\]
$m$ is the minimal integer not smaller than  $t_{max}/T$,
and  $\mathbf L:~\ell_N\to\ell$
is an LTI system.  
\begin{figure}[H]
\setlength{\unitlength}{0.08cm}
\begin{center}\begin{picture}(84,15)(8,64)
\put(24,64){\framebox(16,12){$ \mathbf V$}}
\put(58,64){\framebox(16,12){$ \mathbf L$}}
\put(8,70){\vector(1,0){16}}
\put(40,70){\vector(1,0){18}}
\put(74,70){\vector(1,0){18}}
\put(8,72){$w[n]$}
\put(48,72){$g[n]$}
\put(92,72){$v[n]$}
\end{picture}\end{center}
\caption{Block diagram of the structure of {\bf S}}
\label{fig:compensator}
\end{figure}
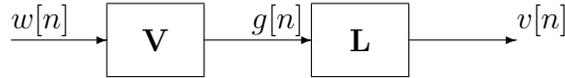
Moreover, $\mathbf L$ can be shown to have a good approximation of the form
$\mathbf L\approx X\mathbf L_0$, where $X$ is a 
static gain matrix, and $\mathbf L_0$
is an LTI model which does not depend on $b_k$ and $t_{k,i}$.
In other words, $\mathbf S$ can be well approximated by combining a Volterra series model
with a short memory, and a {\sl fixed} (long memory) LTI, as long as
the memory depth $t_{max}$ of $\mathbf F$ 
is short, relative to the sampling time $T$.

In most applications, with an appropriate scaling 
and time delay, the system $\mathbf S$
to be inverted can be viewed as a small perturbation of identity, i.e. 
 $\mathbf S=\mathbf I+\mathbf\Delta$. When $\mathbf\Delta$ is  "small" in an appropriate
sense (e.g., has small incremental L2 gain $\|\mathbf\Delta\|\ll1$), the inverse of $\mathbf S$
can be well approximated by 
$\mathbf S^{-1}\approx\mathbf I-\mathbf\Delta=2\mathbf I-\mathbf S$.
Hence the result of this paper suggests a specific structure of the 
compensator (pre-distorter) $\mathbf C\approx\mathbf I-\mathbf\Delta=2\mathbf I-\mathbf S$. 
In other words, a plain Volterra monomials structure is, in general,  not good enough for $\mathbf C$, 
as it lacks the capacity to implement the long-memory
LTI post-filter $\mathbf L$. Instead, $\mathbf C$ should be sought
in the form $\mathbf C=\mathbf I-\mathbf{L_0}X\mathbf{V}$, where $\mathbf V$ is the system
generating all Volterra series monomials of a limited depth and limited degree,
$\mathbf L_0$ is a {\sl fixed} LTI system with a very long time constant, and $X$ is a matrix of coefficients
to be optimized to fit the data available.

\subsection{Ideal Demodulator}

In digital communications literature, demodulation is usually described as 
downconvertion of the passband signal, followed by low-pass filtering (LPF) and 
sampling \cite{Proakis2007}.Often, the low-pass 
filtering (windowing) operation in the transmitter (i.e. in the modulation part of the system) 
is obtain with a filter whose frequency response has significant spectral content outside of band 
of interest (i.e. significant side-lobes are present). In that case the LPF operation after 
downconversion, in demodulator, would null a significant portion of the input signal's spectrum, 
thus introducing additional distortion in the ${\bf DHFM}$ signal chain. Thus distortions introduced 
by the non-ideal demodulation could mask a possibly good preformance of the digital 
predistortion. For that reason, in this paper, we apply demodulation which completely 
recovers the input signal, without introducing additional distortion. We call this operation 
{\sl ideal demodulation}, and in the following derive corresponding mathematical model.

The most commonly known expression for the ideal demodulator inverts not 
$\mathbf M=\mathbf X\mathbf Z$ but $\mathbf M_0=\mathbf X\mathbf H_0\mathbf Z$, i.e., the
modulator which inserts 
$\mathbf H_0$, the {\sl ideal low-pass filter} for the baseband, 
between zero-order hold $\mathbf Z$ and mixer $\mathbf X$,
where $\mathbf H_0$ is the CT LTI system with frequency
response
\[  H_0(\omega)=\begin{cases}
 1,& |\omega|<\omega_b,\\ 
 0,& \text{otherwise}.
\end{cases} \]
Specifically, let
$\mathbf X_c:~\mathcal L\mapsto\mathcal L$ be the {\sl dual
mixer} mapping $x(\cdot)$
to $e(t)=\exp(-j\omega_ct)x(t)$. Let $\mathbf E:~\mathcal L\mapsto\ell$
be the {\sl sampler}, mapping $g(\cdot)$ to $w[n]=g(nT)$. Finally, let $\mathbf A_0$ be the DT
LTI system with frequency response $A_0$ defined by
\[ A_0(\Omega)=P(\Omega/T)^{-1}\quad\text{for}\quad|\Omega|<\pi,\]
where $P$ is the Fourier transform of $p=p(t)$ (\ref{mod_eq}). Then the composition
$\mathbf A_0\mathbf E\mathbf H_0\mathbf X_c\mathbf H\mathbf M_0$ is an identity map.
Equivalently, $\mathbf A_0\mathbf E\mathbf H_0\mathbf X_c$
is the ideal demodulator for $\mathbf M_0$.

For the modulation map $\mathbf M=\mathbf X\mathbf Z$ considered in this paper, the ideal
demodulator has the form $\mathbf A\mathbf E\mathbf H_0\mathbf X_c$, where 
$\mathbf A:~\ell\mapsto\ell$ is the linear system mapping
$w\in\ell(\mathbb C)$ to $s\in\ell(\mathbb C)$ according to
\[  \text{Re}(s) = \mathbf A_{rr}\text{Re}(w)+\mathbf A_{ri}\text{Im}(w),\]
\[  \text{Im}(s) = \mathbf A_{ir}\text{Re}(w)+\mathbf A_{ii}\text{Im}(w),\]
and $\mathbf A_{rr}$, $\mathbf A_{ri}$, $\mathbf A_{ir}$, $\mathbf A_{ii}$ are LTI systems
with frequency responses
$A_{rr}=(P_0-P_i)Q$, $A_{ir}=A_{ri}=-P_qQ$, $A_{ii}=(P_0+P_i)Q$,
where
$Q=(P_0^2-P_i^2-P_q^2)^{-1}$, $P_i=(P^++P^-)/2$, $P_q=(P^+-P^-)/2j$, 
and $P_0,P^+,P^-\in\mathcal L_{2\pi}$ are defined for $|\Omega|<\pi$ by
\[ P_0(\Omega)=P(\Omega/T),~ P^+(\Omega)=P_0(\Omega+\theta),~P^-(\Omega)=P_0(\Omega-\theta)
\]
with $\theta=4\pi M$.

\section{Main Result}

Before stating the main result of this paper, let us introduce some additional 
notation. For $d\in\mathbb N$ and $\tau=(\tau_1,\dots,\tau_d)\in[0,\infty)^d$ 
let $\mathbf F_{\tau}:~\mathcal{L}\to \mathcal{L}$ be the CT system mapping 
inputs $x\in\mathcal L$ to the outputs $y\in\mathcal L$ according to
\[y(t)=x(t-\tau_1)x(t-\tau_2)\dots x(t-\tau_d).\] 
In the rest of this section, many expressions will contain products of the above type, 
where the complex-valued signal $x$ can be written as $x=i+j\cdot q$, with $i$ and 
$q$ representing its real and imaginary part, respectively. It follows that the 
corresponding products would range over delayed real and imaginary parts of $x$. 
As will be shown later (e.g. in (\ref{e_signals})), these product factors can be 
classified into four groups: combinations of delayed or un-delayed, real or imaginary 
part of $x$. This explains appearance of the index set $\{1,2,3,4\}$ which will be used 
to encode these four groups of signals. 

For every ${\bf m}=(m_1,\dots,m_d)\in\{1,2,3,4\}^d$ and integer 
$l\in\{1,2,3,4\}$ let $S_{\bf m}^l$ be the set of all indices $i$ for which $m_i=l$, i.e., 
$S_{\bf m}^l=\{i\in \{1,\dots,d\}:~m_i=l\}$. Furthermore, define 
\[ N_{\bf m}^1=|S_{\bf m}^1\cup S_{\bf m}^2|,\quad N_{\bf m}^2=|S_{\bf m}^3\cup S_{\bf m}^4|.\]
Clearly $N_{\bf m}^1+N_{\bf m}^2=d$ for every ${\bf m}\in\{1,2,3,4\}^d$. 
Let $R_{\bf m}^c = \{-1,1\}^{N_{\bf m}^1}$ and $R_{\bf m}^s = \{-1,1\}^{N_{\bf m}^2}$. 
Let $(\cdot ,\cdot):\mathbb{R}^d\times \mathbb{R}^d \to \mathbb{R}$ 
denote the standard scalar product in $\mathbb{R}^d$. Define the maps 
$\tilde{\sigma},\sigma:\mathbb{R}^d\to \mathbb{R}$ by 
$\tilde{\sigma}(x)=\sum_{i=1}^d x_i$ and $\sigma(x)=\tilde{\sigma}(x)-1$. 
For a given $m\in\{1,2,3,4\}^d$ and $x\in \mathbb{R}^d$ let $\pi_{\bf m}(x)$ be 
the product of all $x_i$ with $i\in S_{\bf m}^3\cup S_{\bf m}^4$. For $i\in\{1,2\}$, 
define projection operators 
$\mathcal{P}_{\bf m}^{i}:\mathbb{R}^d\to\mathbb{R}^{N_{\bf m}^i}$ 
by  
\[\mathcal{P}_{\bf m}^{i}x=
\begin{bmatrix} x_{n_1}&\dots&x_{n_{N_{\bf m}^i}}\end{bmatrix}^T, \quad
\{n_1, \dots, n_{N_{\bf m}^i}\}=S_{\bf m}^{2i-1}\cup S_{\bf m}^{2i},\quad n_1<\dots<n_{N_{\bf m}^i}.\]
The following example should elucidate the above, somewhat involved, 
notation. Let $d=7$ and ${\bf m}=(3,1,4,2,1,3,1)$. Then
\[S_{\bf m}^1=\{2,5,7\}, \quad S_{\bf m}^2=\{4\}, 
\quad S_{\bf m}^3=\{1,6\}, \quad S_{\bf m}^4=\{3\},\]
\[N_{\bf m}^1=|S_{\bf m}^1\cup S_{\bf m}^2|=4, 
\quad N_{\bf m}^2=|S_{\bf m}^3\cup S_{\bf m}^4|=3,\]
\[R_{\bf m}^c=\{-1,1\}^4, \quad R_{\bf m}^2=\{-1,1\}^3,\]
\[\mathcal{P}_{\bf m}^1 x=
\begin{bmatrix} x_2 & x_4 & x_5 & x_7\end{bmatrix}^T, \quad
\mathcal{P}_{\bf m}^2 x=
\begin{bmatrix} x_1 & x_3 & x_6\end{bmatrix}^T,\]
\[\pi_{\bf m}(x)=x_1x_3x_6.\]

Given a vector $\tau\in[0,\infty)^d$ let $\bf k$ be 
the unique vector in $(\mathbb{N}\cup \{0\})^d$
such that $\tau={\bf k}T+\tau'$ and $\tau'\in [0,T)^d$.

Let $\theta:~\mathbb R\to\{0,1\}$ denote the Heaviside step function
$\theta(t)=0$ for $t<0$, $\theta(t)=1$ for $t\ge0$.
For $T\in(0,\infty)$ let $p(t)=p_T(t)= \theta(t)-\theta(t-T)$ denote the basic
pulse shape of the zero-order hold (ZOH) system with sampling time $T$. 
Given $m\in\{1,2,3,4\}^d$ 
and $\tau'\in [0,T)^d$ define 
\[\tau_{min}^{\bf m}=\begin{cases}\max_{i\in S_{\bf m}^2\cup S_{\bf m}^4} \tau'_i,&
|S_{\bf m}^2\cup S_{\bf m}^4|>0,\\0,&\text{otherwise,} \end{cases}\] 
and 
\[\tau_{max}^{\bf m}=\begin{cases} \min_{i\in S_{\bf m}^1\cup S_{\bf m}^3}\tau'_i,&
|S_{\bf m}^1\cup S_{\bf m}^3|>0,\\T,&\text{otherwise.} \end{cases}\]
Let $p_{{\bf m},\tau}:\mathbb{R}\to \mathbb{R}$ be the continuous time signal
 defined by 
\begin{equation}\label{pulse_shape}
p_{{\bf m},\tau}(t)=\begin{cases}\theta(t-\tau_{min}^{\bf m})-\theta(t-\tau_{max}^{\bf m}),& 
\tau_{min}^{\bf m}<\tau_{max}^{\bf m}\\0,&\text{otherwise,} \end{cases}.
\end{equation} 
We denote its Fourier transform by $P_{{\bf m},\tau}(\omega)$.

\begin{figure}[h]
\setlength{\unitlength}{0.1cm}
\begin{center}\begin{picture}(96,11)(2,58)
\put(12,58){\framebox(12,8){$\mathbf M$}}
\put(34,58){\framebox(12,8){$\mathbf  {F_\tau}$}}
\put(56,58){\framebox(12,8){$\mathbf  H$}}
\put(78,58){\framebox(10,8){$\mathbf  D$}}
\put(2,62){\vector(1,0){10}}
\put(24,62){\vector(1,0){10}}
\put(46,62){\vector(1,0){10}}
\put(68,62){\vector(1,0){10}}
\put(88,62){\vector(1,0){10}}
\put(2,64){$w[n]$}
\put(25,64){$x(t)$}
\put(47,64){$y(t)$}
\put(69,64){$z(t)$}
\put(94,64){$v[n]$}
\end{picture}\end{center}
\caption{Block diagram of system $\mathbf{S_\tau= DHF_\tau M}$}
\label{S_tau}
\end{figure}
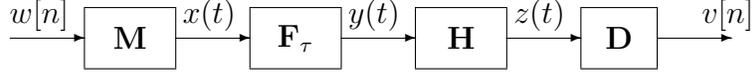

As can be seen from (\ref{ctvs}), the general CT Volterra model is a linear 
combination of subsystems $\mathbf F_{\tau}$, with different $\tau$. Thus, in 
order to establish the desired decomposition $\mathbf {S=LV}$  it is sufficient 
to consider the case $\mathbf{S_\tau=DHF_{\tau}M}$ with a specific $\tau$,
as shown on the block diagram in Fig. \ref{S_tau}. The following theorem gives 
an answer to that question.

\vskip2mm
\noindent\textbf{Theorem 4.1.} 
For $\tau\in[0,\infty)^d$,
the system $\mathbf{DHF_{\tau}M}$
maps $w\in\ell$ to 
\[ v=\mathbf Au\in\ell,\quad\text{with}\quad 
u=\sum_{{\bf m}\in\{1,2,3,4\}^d}x_{\bf m,k}*g_{\bf m},\]
where
\[ i[n]=\text{Re}(w[n]),\quad q[n]=\text{Im}(w[n]),\]
\[x_{\bf m,k}[n]=\prod_{i\in S_{\bf m}^1}i[n-k_i-1] \prod_{i\in S_{\bf m}^2} i[n-k_i]
\prod_{i\in S_{\bf m}^3} q[n-k_i-1] \prod_{i\in S_{\bf m}^4}q[n-k_i],
\end{equation*}
and the sequences (unit sample responses) $g_{\bf m}=g_{\bf m}[n]$ are defined
by their Fourier transforms 
\begin{equation}\label{sasha1}
G_{\bf m}(\Omega)=\frac{ (j)^{N_{\bf m}^2}}{2^d} \sum_{r_c\in R_{\bf m}^c}
\sum_{r_s\in R_{\bf m}^s}{\prod_{l=1}^{N_{\bf m}^2} r_c(l)}\cdot
 P_{{\bf m},\bar\tau}\left(\tilde\Omega\right) \cdot 
e^{-j\omega_c[(r_c,\mathcal{P}_{\bf m}^1\bar\tau) + (r_s,\mathcal{P}_{\bf m}^2\bar\tau)]},
\end{equation}
\[ \tilde\Omega=
\frac{\Omega}{T}-\omega_c \sum_i r_c(i) -\omega_c \sum_l r_s(l)+\omega_c.\]
\vskip3mm

\begin{proof}
We first state and prove the following Lemma, which is a special case of Theorem 4.1, 
in which $\tau$ ranges over $[0,T)^d$ (instead of $\tau\in [0,\infty)^d$), and hence 
${\bf k}=\mathbf{0}$. The proof of Theorem 4.1 follows immediately from this Lemma.

\vskip2mm
\noindent\textbf{Lemma 4.2.} 
The DT system $\mathbf{DHF_{\tau}M}$ with $\tau\in [0,T)^d$
maps  $w\in\ell$ to 
\[ v=\mathbf Au\in\ell,\quad\text{with}\quad 
u=\sum_{{\bf m}\in \{1,2,3,4\}^d}x_{\bf m}*g_{\bf m},\]
where
\[ x_{\bf m}[n]=i[n-1]^{|S_{\bf m}^1|} i[n]^{|S_{\bf m}^2|} 
q[n-1]^{|S_{\bf m}^3|} q[n]^{|S_{\bf m}^4|},\quad
i[n]=\text{Re}(w[n]),\quad q[n]=\text{Im}(w[n]),\]
and the sequences $g_{\bf m}$ are as defined in Theorem 4.1.
\vskip3mm

\begin{proof}
A block diagram of system $\mathbf{DHF_{\tau}M}$ is shown in 
Fig. \ref{S_tau_decomposed}, where $\bf M$ and $\bf D$ are decomposed into 
elementary subsystems as defined in the previous chapters.
In order to prove Lemma 4.1, we first find the analytical expression of signal $y$ as a  
function of $w, \omega_c, T$ and $\tau$, and then we find the map from $y$ to $u$.   

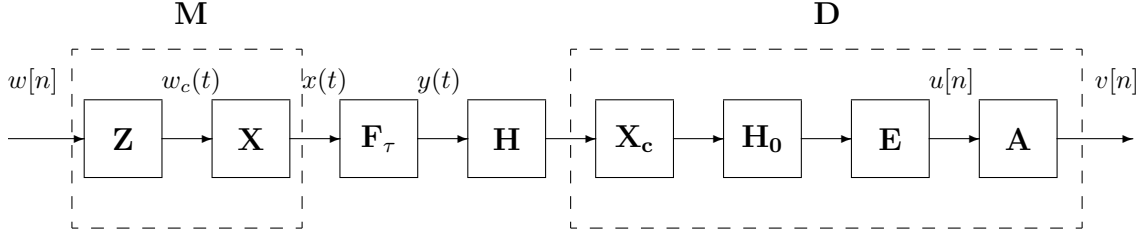
\begin{figure}[h]
\setlength{\unitlength}{0.17cm}
\begin{center}\begin{picture}(91,20)(0,48)
\put(16,52){\framebox(6,6){$\mathbf X$}}
\put(26,52){\framebox(6,6){$ \mathbf{F_\tau}$}}
\put(36,52){\framebox(6,6){$\mathbf H$}}
\put(46,52){\framebox(6,6){$\mathbf{X_c}$}}
\put(56,52){\framebox(6,6){$\mathbf{ H_0}$}}
\put(66,52){\framebox(6,6){$\mathbf E$}}
\put(76,52){\framebox(6,6){$\mathbf A$}}
\put(82,55){\vector(1,0){6}}
\put(0,55){\vector(1,0){6}}
\put(12,55){\vector(1,0){4}}
\put(22,55){\vector(1,0){4}}
\put(32,55){\vector(1,0){4}}
\put(42,55){\vector(1,0){4}}
\put(52,55){\vector(1,0){4}}
\put(62,55){\vector(1,0){4}}
\put(72,55){\vector(1,0){4}}
\put(6,52){\framebox(6,6){$\mathbf Z$}}
\put(44,48){\dashbox(40,14){$ $}}
\put(83,64){$$}
\put(63,64){$\mathbf{D}$}
\put(13,64){$\mathbf M$}
\put(85,59){\footnotesize{$v[n]$}}
\put(0,59){\footnotesize{$w[n]$}}
\put(12,59){\footnotesize{$w_c(t)$}}
\put(23,59){\footnotesize{$x(t)$}}
\put(32,59){\footnotesize{$y(t)$}}
\put(72,59){\footnotesize{$u[n]$}}
\put(5,48){\dashbox(18,14){$ $}}
\end{picture}\end{center}
\caption{Block diagram of system $\mathbf{S_\tau=DHF_{\tau}M}$}
\label{S_tau_decomposed}
\end{figure}

Consider first the case $d=1$ (i.e., $\mathbf{F_{\tau}}$ is just a delay by $\tau\in[0,T)$). 
By definition, the outputs $w_c, x_c$ and $y$ of $\bf Z, X$ and $\bf F$ 
are given by
\begin{equation*}
w_c(t)=\frac1T\sum_{n=-\infty}^{\infty}w[n]p(t-nT)=
\underbrace{\frac1T\sum_{n=-\infty}^{\infty}i[n]p(t-nT)}_{i_c(t)}+
\underbrace{\frac{j}{T}\sum_{n=-\infty}^{\infty}w[n]p(t-nT)}_{jq_c(t)}.
\end{equation*}

\[x(t)=({\bf X}w_c)(t)=\text{Re}\{\exp(j\omega_ct)w_c(t)\},\] 
\begin{equation}\label{y(t)}
y(t)=i_c(t-\tau)\cos(\omega_ct-\omega_c\tau)-q_c(t-\tau)\sin(\omega_ct-\omega_c\tau). 
\end{equation}

\noindent 
Consider the representation $p(t)=p_{1,\tau}(t)+p_{2,\tau}(t)$, where 
\[ p_{1,\tau}(t) =  \theta(t)-\theta(t-\tau),\quad\quad\quad p_{2,\tau}(t) =  \theta(t-\tau)-\theta(t-T).\]
Let ${\bf Z_1}:\ell(\mathbb{C})\to\mathcal L(\mathbb C)$ and 
$\bf Z_2:\ell(\mathbb{C})\to\mathcal L(\mathbb C)$ be the digital-to-analog 
converters with pulse shapes $p_{1,\tau}$ and $p_{2,\tau}$ respectively. 
Let $\bf B$ denote the backshift function mapping $x\in \ell$ to $y={\bf B}x\in\ell$, 
defined by $y[n]=x[n-1]$. Then
\begin{equation}\label{i_and_q_c}
\begin{aligned}
i_c(t-\tau)&=&e_{1,\tau}(t)+e_{2,\tau}(t),\\
q_c(t-\tau)&=&e_{3,\tau}(t)+e_{4,\tau}(t),
\end{aligned}
\end{equation}
where
%
\begin{equation}\label{e_signals}
\begin{aligned}
e_{1,\tau}={\bf Z_1 B}i,\quad\mbox{i.e., } \quad e_{1,\tau}(t) &= \frac1T\sum_{n=-\infty}^\infty i[n-1]p_{1,\tau}(t-nT),\\
e_{2,\tau}={\bf Z_2 }i,\quad\mbox{i.e., } \quad e_{2,\tau}(t) &= \frac1T\sum_{n=-\infty}^\infty i[n]p_{2,\tau}(t-nT),\\
e_{3,\tau}={\bf Z_1 B}q,\quad\mbox{i.e., } \quad e_{3,\tau}(t) &= -\frac1T\sum_{n=-\infty}^\infty q[n-1]p_{1,\tau}(t-nT),\\
e_{4,\tau}={\bf Z_2}q,\quad\mbox{i.e., } \quad e_{4,\tau}(t) &= -\frac1T\sum_{n=-\infty}^\infty q[n]p_{2,\tau}(t-nT).
\end{aligned}
\end{equation}


\noindent It follows from (\ref{y(t)})-(\ref{e_signals}), that the 
output $y(t)$ of $\mathbf{F_{\tau}}$ can be expressed as:
\[y(t) = f_1(t)+f_2(t)+f_3(t)+f_4(t),\] 
where
\begin{equation}\label{signals_f}
f_i(t)=\begin{cases}e_{i,\tau}(t)\cos(\omega_ct-\omega_c\tau),& i=1,2\\
e_{i,\tau}(t)\sin(\omega_ct-\omega_c\tau),& i=3,4 \end{cases}.
\end{equation}
Therefore subsystem $\mathbf{F_{\tau}M}$, mapping $w[n]$ to $y(t)$, can be 
represented as a parallel interconnection of amplitude modulated delayed and 
undelayed in-phase and quadrature components of $w[n]$. This is shown in 
Fig.~\ref{F_tauM}, where  $\bf \tilde D=\bf EH_0X_cH$.

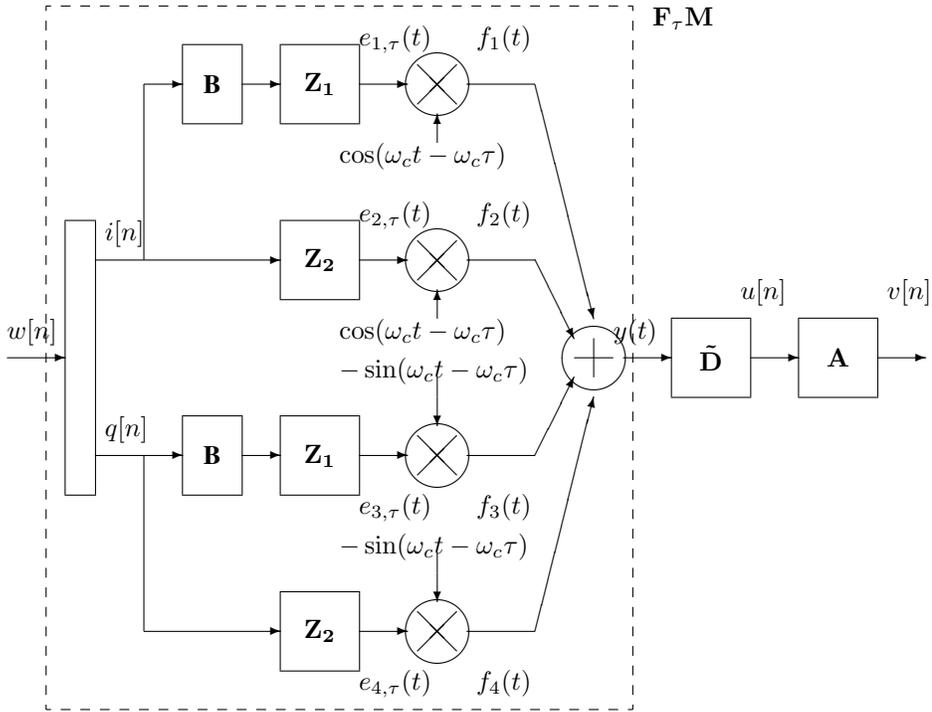
\begin{figure}[H]
\footnotesize
\setlength{\unitlength}{0.13cm}
\begin{center}\begin{picture}(101,72)(0,16)
\put(0,52){\vector(1,0){6}}
\put(0,54){$w[n]$}
\put(6,38){\framebox(3,28){$ $}}
\put(14,62){\line(-1,0){5}}
\put(14,62){\line(0,1){18}}
\put(14,80){\vector(1,0){4}}
\put(18,76){\framebox(6,8){\bf B}}
\put(28,76){\framebox(8,8){$\bf Z_1$}}
\put(9,42){\vector(1,0){9}}
\put(14,42){\line(0,-1){18}}
\put(14,24){\vector(1,0){14}}
\put(28,20){\framebox(8,8){$\bf Z_2$}}
\put(18,38){\framebox(6,8){\bf B}}
\put(24,80){\vector(1,0){4}}
\put(24,42){\vector(1,0){4}}
\put(28,38){\framebox(8,8){$\bf Z_1$}}
\put(14,62){\vector(1,0){14}}
\put(28,58){\framebox(8,8){$\bf Z_2$}}
\put(44,24){\circle{6}}
\put(36,24){\vector(1,0){5}}
\put(36,18){$e_{4,\tau}(t)$}
\put(44,42){\circle{6}}
\put(36,42){\vector(1,0){5}}
\put(44,62){\circle{6}}
\put(36,62){\vector(1,0){5}}
\put(44,80){\circle{6}}
\put(36,80){\vector(1,0){5}}
\put(36,36){$e_{3,\tau}(t)$}
\put(36,66){$e_{2,\tau}(t)$}
\put(36,84){$e_{1,\tau}(t)$}
\put(42,82){\line(1,-1){4}}
\put(42,78){\line(1,1){4}}
\put(42,60){\line(1,1){4}}
\put(42,64){\line(1,-1){4}}
\put(42,40){\line(1,1){4}}
\put(42,44){\line(1,-1){4}}
\put(42,26){\line(1,-1){4}}
\put(42,22){\line(1,1){4}}
\put(44,56){\vector(0,1){3}}
\put(44,74){\vector(0,1){3}}
\put(44,48){\vector(0,-1){3}}
\put(44,30){\vector(0,-1){3}}
\put(34,72){$\cos(\omega_ct-\omega_c\tau)$}
\put(34,54){$\cos(\omega_ct-\omega_c\tau)$}
\put(34,50){$-\sin(\omega_ct-\omega_c\tau)$}
\put(34,32){$-\sin(\omega_ct-\omega_c\tau)$}
\put(48,84){$f_1(t)$}
\put(48,66){$f_2(t)$}
\put(48,36){$f_3(t)$}
\put(48,18){$f_4(t)$}
\put(47,62){\line(1,0){7}}
\put(47,80){\line(1,0){7}}
\put(54,80){\vector(1,-4){6}}
\put(60,52){\circle{6}}
\put(47,42){\line(1,0){7}}
\put(54,42){\vector(1,2){4}}
\put(54,62){\vector(1,-2){4}}
\put(47,24){\line(1,0){7}}
\put(54,24){\vector(1,4){6}}
\put(4,16){\dashbox(60,72){$ $}}
\put(60,54){\line(0,-1){4}}
\put(58,52){\line(1,0){4}}
\put(63,52){\vector(1,0){5}}
\put(68,48){\framebox(8,8){$ \mathbf{\tilde D}$}}
\put(76,52){\vector(1,0){5}}
\put(81,48){\framebox(8,8){$\mathbf A$}}
\put(89,52){\vector(1,0){5}}
\put(62,54){$y(t)$}
\put(75,58){$u[n]$}
\put(90,58){$v[n]$}
\put(66,86){$\mathbf{F_{\tau}M}$}
\put(44,30){\line(0,1){2}}
\put(44,48){\line(0,1){2}}
\put(10,64){$i[n]$}
\put(10,44){$q[n]$}
\end{picture}\end{center}
\caption{Equivalent representation of $\mathbf{DHF_{\tau}M}$}
\label{F_tauM}
\end{figure}

Suppose now that order $d$ of $\mathbf{F_{\tau}}$ is an arbitrary positive 
integer larger than 1, i.e. that 
$\mathbf{F_{\tau}}:x\mapsto x(t-\tau_1)\cdot\dots\cdot x(t-\tau_d)$. 
Then the output $y$ of $\mathbf{F_{\tau}}$ can be represented in the form 
\begin{multline} \label{eq_for_y} 
y(t) = [ i_c(t-\tau_1)\cos(\omega_ct-\omega_c\tau_1)-q_c(t-\tau_1)\sin(\omega_ct-\omega_c\tau_1)]\cdot\\ 
\cdot [ i_c(t-\tau_2)\cos(\omega_ct-\omega_c\tau_2)-q_c(t-\tau_2)\sin(\omega_ct-\omega_c\tau_2)]\cdot\ldots\\
\ldots\cdot [ i_c(t-\tau_d)\cos(\omega_ct-\omega_c\tau_d)-q_c(t-\tau_d)\sin(\omega_ct-\omega_c\tau_d)]. 
\end{multline} 

\noindent 
Let us denote the factors in product in (\ref{eq_for_y}) as $y_i(t)$, i.e. 
\[y_i(t) = i_c(t-\tau_i)\cos(\omega_ct-\omega_c\tau_i) - 
q_c(t-\tau_i)\sin(\omega_ct-\omega_c\tau_i).\] 

\noindent
For each $i$, signal $y_i(t)$ can be represented as the 
output of subsystem $\mathbf F_{\tau_i}\mathbf M$, where $\bf F_\tau$ is 
just a simple delay, as discussed above. 
Therefore system $\mathbf {F_{\tau}M}$, mapping $w$ to $y$, can be 
represented as a parallel interconnection of $d$ subsystems $\mathbf F_{\tau_i}\mathbf M$, 
with corresponding outputs $y_i$, where $y(t)=y_1(t)\cdot\ldots\cdot y_d(t).$ 
This is depicted in Fig.~\ref{F_tauMcompact}. Hence, 
by using the same notation as in Figs~\ref{F_tauM} and~\ref{F_tauMcompact}, 
signal $y(t)$ can be written as 
\begin{equation} 
y(t) = \prod_{i=1}^d y_i(t) = \prod_{i=1}^d (f_1^i(t)+f_2^i(t)+f_3^i(t)+f_4^i(t))
=\sum_{{\bf m}\in [4]^d} f_{m_1}^1(t)\cdot \ldots \cdot f_{m_d}^d(t). 
\end{equation} 

\noindent
We have 
\begin{equation}\label{output_y} 
y(t) = \sum_{{\bf m}\in \{1,2,3,4\}^d} f_{\bf m}(t), 
\end{equation}
where $f_{\bf m}(t)$ denotes the product  $f_{m_1}^1(t)\cdot \ldots \cdot f_{m_d}^d(t)$. 

\noindent
Here componenets $m_i$ of ${\bf m}=(m_1,\ m_2,\ \ldots,\ m_d)\in\{1,2,3,4\}^d$, 
determine which signal $f^i_j$, $j\in\{1,2,3,4\}$ from 
(\ref{signals_f}) participates as a product factor in $f_{\bf m}(t)$. 
With signals $e_{m_i,\tau_i}(t)$ as defined in (\ref{e_signals}), 
it follows that summands in (\ref{output_y}) can be written as 
\begin{equation}\label{product_f} 
f_{\bf m}(t)=(-1)^{N_{\bf m}^2} \prod_{i=1}^d e_{m_i,\tau_i}(t) \cdot
\prod_{k\in S_{\bf m}^1\cup S_{\bf m}^2}\cos(\omega_ct-\omega_c\tau_k) 
\cdot \prod_{l\in S_{\bf m}^3\cup S_{\bf m}^4}\sin(\omega_ct-\omega_c\tau_l). 
\end{equation} 

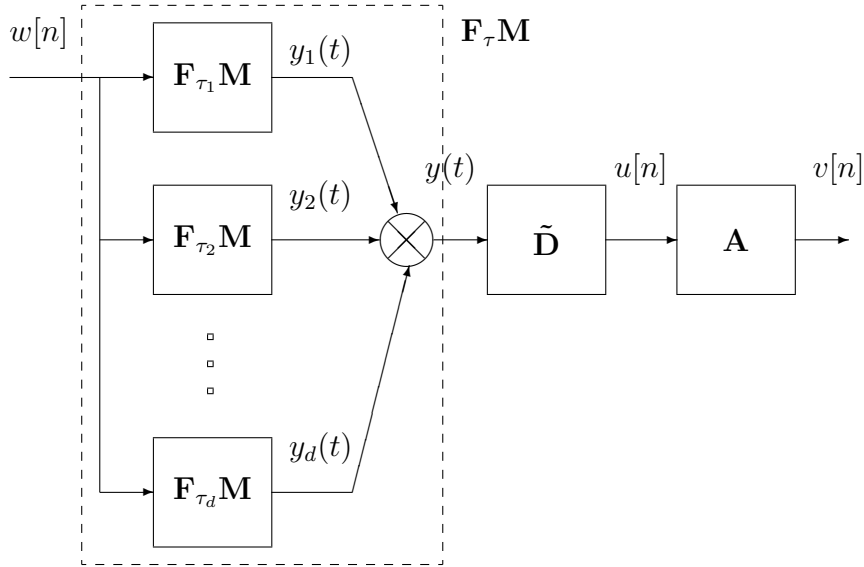
\begin{figure}[h]
\setlength{\unitlength}{0.12cm}
\begin{center}\begin{picture}(100,63)(0,31)
\put(0,86){\vector(1,0){16}}
\put(29,86){\line(1,0){9}}
\put(16,80){\framebox(13,12){$\mathbf F_{\tau_1}\mathbf{M}$}}
\put(16,62){\framebox(13,12){$\mathbf F_{\tau_2}\mathbf{M}$}}
\put(44,68){\circle{6}}
\put(38,40){\vector(1,4){6.3}}
\put(38,86){\vector(1,-3){5}}
\put(29,68){\vector(1,0){12}}
\put(29,40){\line(1,0){9}}
\put(16,34){\framebox(13,12){$\mathbf F_{\tau_d}\mathbf{M}$}}
\put(22,57){\framebox(0.5,0.5)}
\put(22,54){\framebox(0.5,0.5)}
\put(22,51){\framebox(0.5,0.5)}
\put(53,62){\framebox(13,12){$\mathbf {\tilde D}$}}
\put(66,68){\vector(1,0){8}}
\put(74,62){\framebox(13,12){$\mathbf A$}}
\put(87,68){\vector(1,0){6}}
\put(0,90){$w[n]$}
\put(46,75){$y(t)$}
\put(67,75){$u[n]$}
\put(89,75){$v[n]$}
\put(42,66){\line(1,1){4}}
\put(42,70){\line(1,-1){4}}
\put(47,68){\vector(1,0){6}}
\put(31,88){$y_1(t)$}
\put(31,72){$y_2(t)$}
\put(31,44){$y_d(t)$}
\put(10,86){\line(0,-1){46}}
\put(10,40){\vector(1,0){6}}
\put(10,68){\vector(1,0){6}}
\put(8,32){\dashbox(40,62){$ $}}
\put(50,90){$\mathbf{F_{\tau}M}$}
\end{picture}\end{center}
\caption{$\bf F_{\tau}M$ as parallel interconnection of subsystems ${\bf F}_{\tau_i}{\bf M}$}
\label{F_tauMcompact}
\end{figure}

\noindent
Products of cosines and sines in (\ref{product_f}) can be expressed as sums 
of complex exponents as follows
\begin{equation} \label{cosine}
\prod_{k\in S_{\bf m}^1\cup S_{\bf m}^2}\cos(\omega_ct-\omega_c\tau_k)=
 \frac{1}{2^{N_{\bf m}^1}}\sum_{r\in R_{\bf m}^c}e^{j\omega_c\bar \sigma(r)t}
\cdot e^{-j\omega_c(r,\mathcal{P}_{\bf m}^1\tau)},
\end{equation}
\begin{equation} \label{sine}
\prod_{l\in S_{\bf m}^3\cup S_{\bf m}^4}\sin(\omega_ct-\omega_c\tau_l)= 
\frac{1}{(2j)^{N_{\bf m}^2}}\sum_{r\in R_{\bf m}^s}{\prod_{i=1}^{N_{\bf m}^2}r(i)}
\cdot e^{j\omega_c\bar\sigma(r)t}\cdot e^{-j\omega_c(r,\mathcal{P}_m^2\tau)}.
\end{equation}

\noindent 
Recall that the signals $e_{m_i,\tau_i}(t)$ are obtained by applying pulse 
amplitude modulation with pulse signals $p_{1,\tau_i}(t)$ or $p_{2,\tau_i}(t)$ 
on in-phase or quadrature components $i$ and $q$ of the input signal (or their 
delayed counterparts ${\bf B}i$ and ${\bf B}q$). Let $e_{{\bf m},\tau}(t)$ be the 
product of signals $e_{m_i,\tau_i}(t)$ (as given in (\ref{product_f})). 
We now 
derive an expression for $e_{{\bf m},\tau}(t)$ as a function of signals $i, q, {\bf B}i$ 
and ${\bf B}q$. We first investigate signal $e_{{\bf m},\tau}(t)$ for $t\in[nT,(n+1)T)$, 
with $n>1$ an integer. There are three possible cases:

\begin{itemize}

\item [(i)]  $S_{\bf m}^2 \cup S_{\bf m}^4 = \emptyset$, i.e.  
signals $e_{m_i,\tau_i}(t)$ were all obtained by applying 
pulse amplitude modulation with $p_{1,\tau}(t)$.
It immediately follows that product $e_{{\bf m},\tau}(t)$ of signals 
$e_{m_i,\tau_i}(t)$ is nonzero only for $t\in[nT,nT+\tau_{max})$ , 
where $\tau_{max}=\min_{i}\tau_i$.

\item [(ii)] $S_{\bf m}^1 \cup S_{\bf m}^3 = \emptyset$, i.e. 
signals $e_{m_i,\tau_i}(t)$ were all obtained by applying 
pulse amplitude modulation with $p_{2,\tau}(t)$.
It immediately follows that product $e_{{\bf m},\tau}(t)$ of signals 
$e_{m_i,\tau_i}(t)$ is nonzero only for $t\in[nT+\tau_{min},(n+1)T)$, 
where $\tau_{min}=\max_{i}\tau_i$.

\item [(iii)] Both $S_{\bf m}^1\cup S_{\bf m}^3$ and $S_{\bf m}^2\cup S_{\bf m}^4$ 
are non-empty. Let $\tau_{min}=\max_{i\in S_{\bf m}^2\cup S_{\bf m}^4}\tau_i$ 
and $\tau_{max}=\min_{i\in S_{\bf m}^1\cup S_{\bf m}^3}\tau_i$. It follows that 
$e_{{\bf m},\tau}(t)=0$ for all $t\in[nT,(n+1)T)$ if $\tau_{min}>\tau_{max}$.
Otherwise it is nonzero for 
$t\in[nT+\tau_{min},nT+\tau_{max})$. This is depicted in 
Fig. \ref{delays} (for the sake of simplicity, only in-phase component 
$i$ is considered, but in general signals $q$ and ${\bf B}q$ 
would appear too).
\end{itemize}

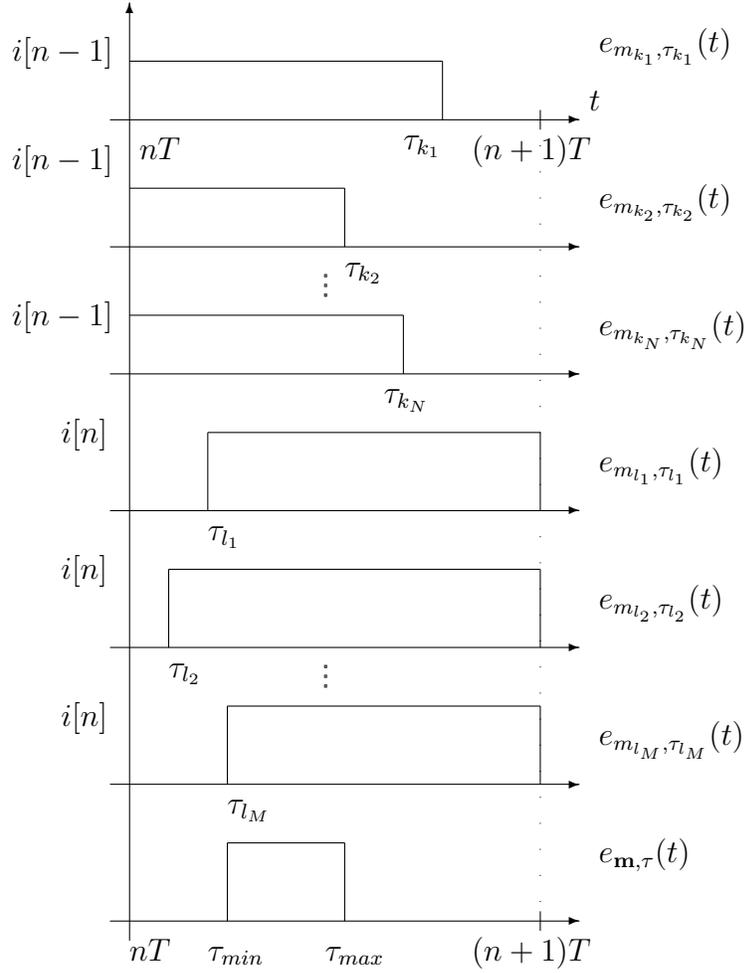
\begin{figure}[h]
\setlength{\unitlength}{0.13cm}
\begin{center}\begin{picture}(64,100)(16,-4)
\put(20,2){\vector(1,0){48}}
\put(20,30){\vector(1,0){48}}
\put(20,44){\vector(1,0){48}}
\put(20,71){\vector(1,0){48}}
\put(20,84){\vector(1,0){48}}
\put(22,0){\vector(0,1){96}}
\qbezier[30](64,84)(64,44)(64,4)
\put(30,44){\line(0,1){8}}
\put(30,52){\line(1,0){34}}
\put(64,52){\line(0,-1){8}}

\put(26,38){\line(1,0){38}}
\put(64,38){\line(0,-1){8}}
\put(26,38){\line(0,-1){8}}

\put(30,41){$\tau_{l_1}$}
\put(26,27){$\tau_{l_2}$}
\put(32,13){$\tau_{l_M}$}
\put(42,28){\framebox(0.1,0.1){$ $}}
\put(42,27){\framebox(0.1,0.1){$ $}}
\put(42,26){\framebox(0.1,0.1){$ $}}
\put(22,90){\line(1,0){32}}
\put(54,90){\line(0,-1){6}}
\put(22,77){\line(1,0){22}}
\put(44,77){\line(0,-1){6}}
\put(42,68){\framebox(0.1,0.1){$ $}}
\put(42,67){\framebox(0.1,0.1){$ $}}
\put(42,66){\framebox(0.1,0.1){$ $}}
\put(20,58){\vector(1,0){48}}
\put(22,64){\line(1,0){28}}
\put(50,64){\line(0,-1){6}}
\put(50,81){$\tau_{k_1}$}
\put(44,68){$\tau_{k_2}$}
\put(48,55){$\tau_{k_N}$}
\put(32,2){\line(0,1){8}}
\put(32,10){\line(1,0){12}}
\put(44,10){\line(0,-1){8}}
\put(70,91){$e_{m_{k_1},\tau_{k_1}}(t)$}
\put(70,75){$e_{m_{k_2},\tau_{k_2}}(t)$}
\put(70,62){$e_{m_{k_N},\tau_{k_N}}(t)$}
\put(70,48){$e_{m_{l_1},\tau_{l_1}}(t)$}
\put(70,34){$e_{m_{l_2},\tau_{l_2}}(t)$}
\put(32,16){\line(0,1){8}}
\put(32,24){\line(1,0){32}}
\put(64,24){\line(0,-1){8}}
\put(20,16){\vector(1,0){48}}
\put(70,20){$e_{m_{l_M},\tau_{l_M}}(t)$}
\put(70,8){$e_{{\bf m},\tau}(t)$}
\put(30,-2){$\tau_{min}$}
\put(42,-2){$\tau_{max}$}
\put(22,-2){$nT$}
\put(57,-2){$(n+1)T$}
\put(64,1){\line(0,1){2}}
\put(69,85){$t$}
\put(15,51){$i[n]$}
\put(15,37){$i[n]$}
\put(15,22){$i[n]$}
\put(10,90){$i[n-1]$}
\put(10,79){$i[n-1]$}
\put(10,63){$i[n-1]$}
\put(23,80){$nT$}
\put(57,80){$(n+1)T$}
\put(64,83){\line(0,1){2}}
\end{picture}\end{center}
\caption{Signal $e_{{\bf m},\tau}$ for $S_{\bf m}^1\cup S_{\bf m}^3=\{k_1,k_2,\dots,k_N\}$ 
and $S_{\bf m}^2\cup S_{\bf m}^4=\{l_1,l_2,\dots,l_M\}$, where $N+M=d$}
\label{delays}
\end{figure}

\noindent 
The above discussion implies that the signal $e_{{\bf m},\tau}(t)$ 
can be expressed as
\begin{equation}\label{e_m_tau}
e_{{\bf m},\tau}(t) = \sum_{n=-\infty}^\infty 
x_{\bf m}[n]p_{{\bf m},\tau}(t-nT),
\end{equation}
\noindent 
where $p_{{\bf m},\tau}(t)$ was defined in (\ref{pulse_shape}), and 
DT signal $x_{\bf m}=x_{\bf m}[n]$ is defined as
\[ x_{\bf m}[n] = i[n]^{|S_{\bf m}^1|}\cdot i[n-1]^{|S_{\bf m}^2|}
\cdot q[n]^{|S_{\bf m}^3|}\cdot q[n-1]^{|S_{\bf m}^4|}. \] 

\noindent
From (\ref{product_f})-(\ref{e_m_tau}), it follows that $f_{\bf m}(t)$ 
can be written as
\begin{equation} \label{f_m}
f_{\bf m}(t) = f_{m_1}^1(t)\cdot \ldots \cdot f_{m_d}^d(t) = 
\left(\sum_{r_c\in R_{\bf m}^c} \sum_{r_s\in R_{\bf m}^s}  
C_{r_c,r_s} \cdot e^{j \sigma(r_c,r_s)\omega_ct}\right)
\sum_{n=-\infty}^\infty x_{\bf m}[n]p_{{\bf m},\tau}(t-nT),
\end{equation}

\noindent
where $\sigma([r_c^T,r_s^T]^T) =\sigma(r) = \sum_k r_c(k) + \sum_l r_s(l)$, and 
\begin{equation} \label{Crcrs}
C_{r_c,r_s}=\frac{ (j)^{N_{\bf m}^2}}{2^d}\cdot 
e^{-j\omega_c[(r_c,\mathcal{P}_{\bf m}^1\tau) 
+ (r_s,\mathcal{P}_{\bf m}^2\tau)]}
\cdot \prod_{l=1}^{N_{\bf m}^2} r_s(l),
\end{equation}

\noindent
depends only on ${\bf m}$. 
Therefore, the output signal $y$ of system $\bf F_\tau M$, can be 
expressed in terms of $w$ (more precisely in terms of $i$ and $q$) by 
plugging the expression (\ref{f_m}) for $f_{\bf m}(t)$ into 
(\ref{output_y}).
Thus we have found an explicit input-output relationship of system 
$\bf F_\tau M$, which concludes the first part of the proof. 

\vskip3mm
In order to find the relationship between input and output signals of the 
subsystem $\bf DH$, i.e. $y$ and $v$, respectively, we first express 
signal $u$ as a function of $y$ (see Fig. \ref{S_tau_decomposed}). 
Recall that $u={\bf\tilde D}y = {\bf EH_0X_cH}y$. Let $U(\Omega)$ and 
$Y(\omega)$ denote the Fourier transforms of signals $u[n]$ and $y(t)$ 
respectively. 
Also let $H(\omega)$ and $H_0(\omega)$ be the frequency responses 
of ideal band-pass and low-pass filters $\bf H$ and $\bf H_0$, given by
\begin{equation}
\begin{aligned}
H(\omega) &= \begin{cases} 1, & \omega_c-\pi/T \le |\omega| \le \omega_c+\pi/T  \\
0, & \text{o/w} \end{cases},\\
H_0(\omega) &= \begin{cases} 1, & |\omega| \le \pi/T  \\
0, & \text{o/w} \end{cases}.
\end{aligned}
\end{equation}

\noindent
The following sequence of equalities holds
\begin{align*}
\mathcal F\{{\bf H}y\} &= Y(\omega)H(\omega),\\
\mathcal F\{{\bf X_cH}y\} &= Y(\omega+ \omega_c)H(\omega+ \omega_c),\\
\mathcal F\{{\bf H_0X_cH}y\} &= Y(\omega+ \omega_c)H(\omega+ \omega_c)H_o(\omega),\\
U(\Omega) &= Y \left( \frac{\Omega}T+ \omega_c \right) 
H \left( \frac{\Omega}T+ \omega_c \right) H_0 \left( \frac{\Omega}T \right).
\end{align*}

\noindent
From the definition of $H(\omega)$ and $H_0(\omega)$, $U(\Omega)$ simplifies to
\begin{equation} \label{U}
U(\Omega) = Y \left( \frac{\Omega}T+ \omega_c \right).
\end{equation}
Equation (\ref{U}) gives frequency domain relationship between $y$ and $u$. 
 
Next we express $Y(\omega)$ in terms of 
$X_{\bf m}(\Omega)=\mathcal F\{x_{\bf m}[n]\}$. For the sake of simplicity, 
we assume that $y(t)$ is equal to just one signal $f_{\bf m}(t)$ for some 
fixed ${\bf m}$, i.e. we omit the sum in (\ref{output_y}). 
It follows from (\ref{f_m}) that
\begin{equation*}
\mathcal F\{f_{\bf m}(t)\} =Y(\omega) = 
\sum_{r_c\in R_{\bf m}^c}\sum_{r_s\in R_{\bf m}^s}C_{r_c,r_s}
X_{\bf m} \left( \omega T - \sigma(r)\cdot \omega_cT \right) 
P_{{\bf m},\tau} \left( \omega - \sigma(r)\cdot \omega_c \right).
\end{equation*}

\noindent
Since $\sigma(r)\in\mathbb{Z}$ and $\omega_cT =2\pi n$, where 
$n \in \mathbb{Z}$, we get 
\begin{equation}\label{Y}
Y(\omega) = X_{\bf m}(\omega T) \cdot 
\sum_{r_c \in R_{\bf m}^c}\sum_{r_s\in R_{\bf m}^s}C_{r_c,r_s}
P_{{\bf m},\tau} \left( \omega - \sigma(r)\cdot \omega_c \right).
\end{equation}

\noindent
It follows from (\ref{U}) and (\ref{Y}) that 
\[U(\Omega) = X_{\bf m}(\Omega) \cdot 
\sum_{r_c \in R_{\bf m}^c}\sum_{r_s\in R_{\bf m}^s}C_{r_c,r_s}
P_{{\bf m},\tau}\left( \frac{\Omega}T - \omega_c\cdot \sigma(r) + \omega_c\right),\]
with $C_{r_c,r_s}$ as given in (\ref{Crcrs}).

\noindent
Therefore, the frequency response $G_{\bf m}(\Omega)$ of a LTI system 
mapping $x_{\bf m}$ to $u$ is given by
\begin{equation*} 
G_{\bf m}(\Omega) =   \sum_{r_c\in R_{\bf m}^c}\sum_{r_s\in R_{\bf m}^s}C_{r_c,r_s}
P_{{\bf m},\tau}\left(\frac{\Omega}T - \omega_c\cdot \sigma(r) + \omega_c\right).
\end{equation*}
%
This concludes the proof of Lemma 4.2.
\end{proof}

\vskip2mm
In Lemma 4.2, it was assumed that $\tau_i\in[0,T),\ \forall i\in[d]$, 
but in general $\tau_i$ can take any positive real value depending 
on the depth of (2), i.e. vector $\bf k$ associated with $\tau$ is not 
necessarily zero vector. Suppose now that $\tau = {\bf k}T+\bar {\tau}$, 
where $\bar{\tau}\in[0,T)^d$, and $\bf k\not= 0$. In the rest 
of this proof we adopt the same notation for corresponding signals and 
systems as in the proof of Lemma 4.2. 

Clearly, mapping from $y$ to $u$ is identical to the one derived for 
$\tau\in [0,T)$. Thus, in order to prove the statement of Theorem 4.1 
we only have to find relationship between $w$ and $y$. Let $d=1$, i.e. 
$\tau = kT + \bar\tau$, with $k\in \mathbb{N}$ and $\bar\tau\in[0,T)$. 
Analogously to the case in the proof of Lemma 4.2, it follows that signal 
$y$ can be expressed as 
\[y(t)=[e_{1,\tau}(t)+e_{2,\tau}(t)]\cos(\omega_c t-\omega_c\tau)+
[e_{3,\tau}(t)+e_{4,\tau}(t)]\sin(\omega_c t-\omega_c\tau),\]
where
\begin{equation}\label{e_signals_new}
\begin{aligned}
e_{1,\tau}={\bf Z_1 B}^{k+1}i,\quad\mbox{i.e., } \quad 
e_{1,\tau}(t) &= \frac1T\sum_{n=-\infty}^\infty i[n-k-1]p_{1,\bar\tau}(t-nT),\\
e_{2,\tau}={\bf Z_2 B}^ki,\quad\mbox{i.e., } \quad 
e_{2,\tau}(t) &= \frac1T\sum_{n=-\infty}^\infty i[n-k]p_{2,\bar\tau}(t-nT),\\
e_{3,\tau}={\bf Z_1 B}^{k+1}q,\quad\mbox{i.e., } \quad 
e_{3,\tau}(t) &= -\frac1T\sum_{n=-\infty}^\infty q[n-k-1]p_{1,\bar\tau}(t-nT),\\
e_{4,\tau}={\bf Z_2 B}^k q,\quad\mbox{i.e., } \quad 
e_{4,\tau}(t) &= -\frac1T\sum_{n=-\infty}^\infty q[n-k]p_{2,\bar\tau}(t-nT).
\end{aligned}
\end{equation}
Here ${\bf B}^k$ denotes the composition of $\bf B$ with itself $k$ times, 
i.e. ${\bf B}^k:x[n]\mapsto y[n]=x[n-k]$. 

For $d>1$, reasoning similar to that in the proof of Lemma 4.2, leads to the 
following expression for $e_{{\bf m},\tau}$:
\begin{equation}\label{e_m_tau_new}
e_{{\bf m},\tau}(t) = \sum_{n=-\infty}^\infty x_{\bf m, k}[n]p_{{\bf m},\bar\tau}(t-nT),
\end{equation}
where
\begin{equation*}
x_{\bf m,k}[n]=\prod_{i\in S_{\bf m}^1}i[n-k_i-1] \cdot \prod_{i\in S_{\bf m}^2} i[n-k_i]
\cdot \prod_{i\in S_{\bf m}^3} q[n-k_i-1] \cdot \prod_{i\in S_{\bf m}^4}q[n-k_i],
\end{equation*}
and $p_{{\bf m},\bar\tau}(t)$ is defined in (\ref{pulse_shape}). 
Let $X_{\bf m, k}=X_{\bf m, k}(\Omega)$ be the Fourier transform of 
$x_{\bf m,k}$. With (\ref{e_m_tau_new}) at hand, it is straightforward 
to find the analytic expression for $U=\mathcal Fu$, in terms of 
$X_{\bf m, k}$. Similarly to (\ref{f_m})-(\ref{Y}), the Fourier transform 
$Y=\mathcal{F}y$, can be written as
\begin{equation}\label{Y_new}
Y(\omega) = X_{\bf m,k}(\omega T) \cdot 
\sum_{r_c \in R_{\bf m}^c}\sum_{r_s\in R_{\bf m}^s}C_{r_c,r_s}
P_{{\bf m},\bar\tau} \left( \omega - \sigma(r)\cdot \omega_c \right),
\end{equation}
where 
\begin{equation} \label{Crcrs_new}
C_{r_c,r_s}=\frac{ (j)^{N_{\bf m}^2}}{2^d}\cdot 
e^{-j\omega_c[(r_c,\mathcal{P}_{\bf m}^1\bar\tau) 
+ (r_s,\mathcal{P}_{\bf m}^2\bar\tau)]}
\cdot \prod_{l=1}^{N_{\bf m}^2} r_s(l),
\end{equation}
It follows from (\ref{U}) and (\ref{Y_new}) that 
\[U(\Omega) = X_{\bf m,k}(\Omega) \cdot \sum_{r_c \in R_{\bf m}^c}
\sum_{r_s\in R_{\bf m}^s}C_{r_c,r_s} P_{{\bf m},\bar\tau}
\left( \frac{\Omega}T - \omega_c\cdot \sigma(r) + \omega_c\right).\]
Statement of the Theorem now imediatelly proceeds from the above 
equality. This concludes the proof.

\end{proof}
Block diagram of system $\bf S_\tau=DHF_\tau M$, as suggested in 
the statement of Theorem 4.1, is shown in Fig. \ref{system_S_final}. 
System $\bf S_\tau$ can be represented as a parallel interconnection 
of DT nonlinear Volterra subsystems $\bf V_{m,k}$ mapping input signal 
$w$ into output signal $x_{\bf m,k}$, and DT LTI systems $\bf G_{m,k}$ 
mapping input signal $x_{\bf m,k}$ into output signal $u_{\bf m}$.


\section{Discussion}
\subsection{Effects of oversampling}
The analytical result of this paper suggests a special structure 
of a digital pre-distortion compensator which appears to be, in first 
approximation, both necessary and sufficient to match the discrete time 
dynamics resulting from combining modulation and demodulation with a 
dynamic non-linearity in continuous time. The "necessity" somewhat 
relies on the input signal $u$ having "full" spectrum. 
In digital communications it is very common practice to oversample baseband signal 
(symbols), and shape its spectrum (samples), before it is modulated onto a carrier 
\cite{Proakis2007}.
In the case of large oversampling ratios, from symbol to sample space, 
the effective band of the signal containing symbol information is 
small compared to the band assigned by the regulatory agency. 
So in order to transmit symbol information without distortion, the 
reconstruction filter has to match the frequency response of the ideal baseband 
model LTI filter only on this effective band (and the rest can be zeroed-out by 
applying a smoothing filter after demodulation). This now allows for 
reconstruction filters in baseband equivalent model to be not just smooth, but also 
continuous, and thus well approximable by short memory FIR filters. 
This in turn implies that the plain Volterra structure with relatively short memory can capture 
dynamics of such system well enough, possibly diminishing the need for any special 
models.
While, theoretically, the baseband signal $u$ is supposed to be shaped so that only a
lower DT frequency spectrum of it remains significant (i.e. oversampling is employed), 
a practical implementation of amplitude-phase modulation will frequently employ a  
signal component separation approach, such as LINC \cite{Cox1974}, where the
low-pass signal $u$ is decomposed into two components of constant amplitude,
$u=u_1+u_2$, $|u_1[n]|\equiv|u_2[n]|=\text{const}$, after which the components
$u_i$ are fed into two separate modulators, to produce continuous time
outputs $y_1,y_2$, to be combined into a single output $y=y_1+y_2$. Even when
$u$ is band-limited, the resulting components $u_1,u_2$ are not, and the full range of
modulator's nonlinearity is likely to be engaged when producing $y_1$ and $y_2$. 
Also in high-speed wideband communication systems, the
oversampling ratio is usually limited by the speed that the digital baseband and DAC
are able to sustain, therefore the latter scenario described is usually encountered and
the compensator model should be able to take care of this factor.

\subsection{Extension to OFDM}
Orthogonal frequency-division multiplexing (OFDM) is a multicarrier digital modulation 
scheme that has been the dominant technology for broadband multicarrier communications 
in the last decade. 
Compared with single-carrier digital modulation, by increasing the effective symbol 
length and employing many carriers for transmission, OFDM theoretically eliminates 
the problem of multi-path channel fading, 
which is the main type of disturbance on a terrestrial transmission 
path. It also mitigates low spectrum efficiency, impulse noise, and frequency selective 
fading \cite{Proakis2007,Goldsmith2005}. 
One of the major drawbacks of OFDM is the relatively large Peak-to-Average Power 
Ratio (PAPR) \cite{Han2005}. This makes OFDM very 
sensitive to the nonlinear distortion introduced by high PA, which causes 
in-band as well as out-of-band (i.e. adjacent channel) 
radiation, decreasing spectral efficiency \cite{Shi1996}. For that reason 
linearization techniques play very important role in OFDM, and have been studied 
extensively \cite{OFDM DPD1}-\cite{OFDM DPD3}.

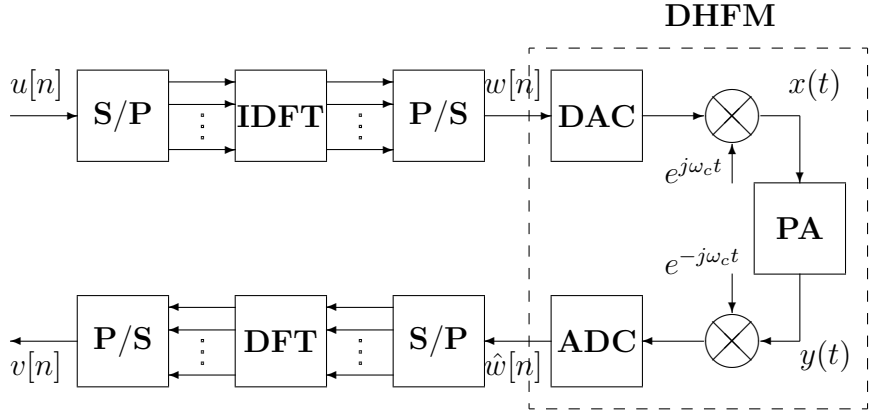
\begin{figure}[h]
\setlength{\unitlength}{0.15cm}
\begin{center}\begin{picture}(76,41)(0,33)
\put(0,66){\vector(1,0){6}}
\put(6,62){\framebox(8,8){$\bf S/P$}}
\put(14,69){\vector(1,0){6}}
\put(14,67){\vector(1,0){6}}
\put(17,66){\framebox(0.1,0.1)}
\put(17,65){\framebox(0.1,0.1)}
\put(17,64){\framebox(0.1,0.1)}
\put(14,63){\vector(1,0){6}}
\put(20,62){\framebox(8,8){$\bf IDFT$}}
\put(28,69){\vector(1,0){6}}
\put(28,67){\vector(1,0){6}}
\put(31,66){\framebox(0.1,0.1)}
\put(31,65){\framebox(0.1,0.1)}
\put(31,64){\framebox(0.1,0.1)}
\put(28,63){\vector(1,0){6}}
\put(34,62){\framebox(8,8){$\bf P/S$}}
\put(42,66){\vector(1,0){6}}
\put(48,62){\framebox(8,8){$\bf DAC$}}
\put(56,66){\vector(1,0){5.5}}
\put(64,66){\circle{5}}
\put(62.3,64.3){\line(1,1){3.4}}
\put(62.3,67.7){\line(1,-1){3.4}}
\put(70,66){\vector(0,-1){6}}
\put(66,52){\framebox(8,8){$\bf PA$}}
\put(70,46){\vector(-1,0){3.5}}
\put(64,46){\circle{5}}
\put(62.3,44.3){\line(1,1){3.4}}
\put(62.3,47.7){\line(1,-1){3.4}}
\put(61.5,46){\vector(-1,0){5.5}}
\put(48,42){\framebox(8,8){$\bf ADC$}}
\put(48,46){\vector(-1,0){6}}
\put(34,42){\framebox(8,8){$\bf S/P$}}
\put(34,49){\vector(-1,0){6}}
\put(34,47){\vector(-1,0){6}}
\put(31,46){\framebox(0.1,0.1)}
\put(31,45){\framebox(0.1,0.1)}
\put(31,44){\framebox(0.1,0.1)}
\put(34,43){\vector(-1,0){6}}
\put(20,42){\framebox(8,8){$\bf DFT$}}
\put(20,49){\vector(-1,0){6}}
\put(20,47){\vector(-1,0){6}}
\put(17,46){\framebox(0.1,0.1)}
\put(17,45){\framebox(0.1,0.1)}
\put(17,44){\framebox(0.1,0.1)}
\put(20,43){\vector(-1,0){6}}
\put(6,42){\framebox(8,8){$\bf P/S$}} 
\put(6,46){\vector(-1,0){6}}
\put(46,40){\dashbox(30,32){$ $}}
\put(58,74){$\bf DHFM$}
\put(0,68){$u[n]$}
\put(69,68){$x(t)$}
\put(70,44){$y(t)$}
\put(0,43){$v[n]$}
\put(42,68){$w[n]$}
\put(42,43){$\hat w[n]$}
\put(70,52){\line(0,-1){6}}
\put(66.5,66){\line(1,0){3.5}}
\put(64,60){\vector(0,1){3.5}}
\put(64,52){\vector(0,-1){3.5}}
\put(58,60){$e^{j\omega_ct}$}
\put(58,52){$e^{-j\omega_ct}$}
\end{picture}\end{center}
\caption{Block diagram of a typical implementation of OFDM}
\label{OFDM}
\end{figure}

Fig. \ref{OFDM} shows a block diagram of the typical implementation 
of an $N$-carrier OFDM system. Input stream of symbols $u[n]$, with bandwidth $B$,  
is first converted into blocks of lenght $N$ by serial-to-parallel conversion, 
which are then fed to an $N$-point inverse FFT block. Output of this block is 
then transformed with a parallel-to-serial converter into a stream of $N$ 
samples $v[k]$, with bandwidth $B$ (usually this bandwidth is larger than 
the input symbols' bandwidth, but in our discussion 
we ignore introduction of the guard interval (i.e. addition of cyclic 
redundancy), which is usually used to mitigate the impairments of the multipath 
radio channel, as it does not affect aplicability of the baseband model and the DPD 
proposed in this paper). Digital-to-analog convertion is then applied to $w[k]$, 
and its output is used to modulate a single carrier. As can be seen from 
Figure \ref{OFDM}, sequence $w[k]$ can be seen as 
an input to a system which can be modeled as the $\bf DHFM$ system investigated 
in the previous chapter. In our derivation of the baseband model, choice of the 
input symbols' values (e.g. QPSK, QAM, etc.), was not relevant to the actual 
derivation. In other words, input symbols can take any value from $\mathbb C$, 
hence sequence $w[k]$ can be considered as a legitimate input sequence to a system 
modeled as $\bf DHFM$. This suggests that our baseband model, and its 
corresponding DPD structure, can be possibly used for distortion reduction 
in OFDM modulation applications. \\

\section{Simulation Results}
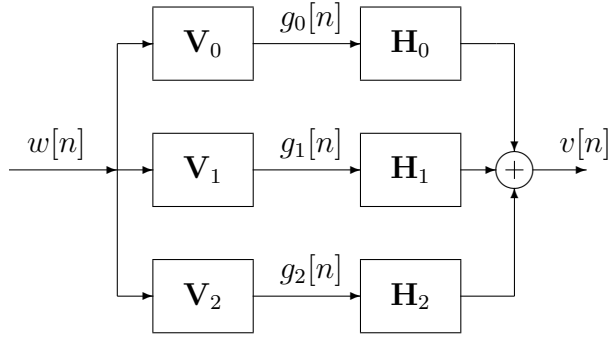
\begin{figure}[h]
\setlength{\unitlength}{0.12cm}
\begin{center}\begin{picture}(64,36)(8,36)
\put(24,64){\framebox(11,8){$\mathbf V_0$}}
\put(24,50){\framebox(11,8){$\mathbf V_1$}}
\put(24,36){\framebox(11,8){$\mathbf V_2$}}
\put(35,68){\vector(1,0){12}}
\put(47,64){\framebox(11,8){$\mathbf H_0$}}
\put(35,54){\vector(1,0){12}}
\put(47,50){\framebox(11,8){$\mathbf H_1$}}
\put(35,40){\vector(1,0){12}}
\put(47,36){\framebox(11,8){$\mathbf H_2$}}
\put(20,68){\vector(1,0){4}}
\put(20,68){\line(0,-1){28}}
\put(20,40){\vector(1,0){4}}
\put(20,54){\vector(1,0){4}}
\put(8,54){\vector(1,0){12}}
\put(58,54){\vector(1,0){4}}
\put(64,54){\circle{4}}
\put(58,68){\line(1,0){4}}
\put(62,68){\line(1,0){2}}
\put(64,68){\vector(0,-1){12}}
\put(58,40){\line(1,0){6}}
\put(64,40){\vector(0,1){12}}
\put(66,54){\vector(1,0){6}}
\put(64,55){\line(0,-1){2}}
\put(63,54){\line(1,0){2}}
\put(10,56){$w[n]$}
\put(38,56){$g_1[n]$}
\put(38,70){$g_0[n]$}
\put(38,42){$g_2[n]$}
\put(69,56){$v[n]$}
\end{picture}\end{center}
\caption{Proposed compensator structure}
\label{fig:Compensator_structure}
\end{figure}

In this section, aided by MATLAB simulations, we illustrate performance of the 
proposed compensator structure. We compare this structure with some standard 
compensator structures, together with the ideal compensator, and show 
that it closely resembles dynamics of ideal compensator, thus achieving near optimal 
compensation performance. \\
The underlying system {\bf S} is shown in Figure~\ref{fig:whole_chain}, with the 
analog channel subsystem {\bf F} given by
\begin{equation} \label{eq:distortion}
(Fx)(t)=x(t)-\delta\cdot x(t-\tau_1)x(t-\tau_2)x(t-\tau_3),
\end{equation}
where $0\le \tau_1 \le \tau_2 \le \tau_3 \le T$, with $T$ sampling time, and 
$\delta>0$ parameter specifying magnitude of distortion $\bf \Delta$ in 
${\bf S}={\bf I}+{\bf \Delta}$. We assume that parameter $\delta$ is relatively 
small, in particular $\delta \in (0,0.2)$, so that the inverse $\mathbf S^{-1}$ of {\bf S} can be well 
approximated by $2\mathbf I - \mathbf S$. 
Then our goal is to build compensator $\mathbf C=\mathbf S^{-1}$ with 
different structures, and compare their performance, which is measured as 
output Error Vector Magnitude (EVM)  \cite{Vuolevi2003} defined, for a given
input-output pair $(u,\hat u)$, as
\[\textrm{EVM(dB)} =20\log_{10} \left(\frac{||u-\hat u||_2}{||u||_2}\right).\]
Analytical results from the previous section suggest that the compensator structure 
should be of the form depicted in Figure~\ref{fig:compensator}.
It is easy to see from the proof of Theorem 4.1, that transfer functions in {\bf L}, 
from each nonlinear component $g_k[n]$ of $g[n]$, to the output $v[n]$, are 
smooth functions, hence can be well approximated by low order 
polynomials in $\Omega$. In this example we choose second order polynomial 
approximation of components of {\bf L}. This observation, together with the true 
structure of {\bf S}, suggests that compensator {\bf C} should be fit
within a family of models with structure shown on the block diagram in 
Fig~\ref{fig:Compensator_structure}, where

\begin{itemize}
\item [(a)] Subsystems $\mathbf H_i, i=1,2,3$, are LTI systems, with 
transfer functions $H_i$ given by
\[H_0(e^{j\Omega})=1, H_1(e^{j\Omega})=j\Omega, 
H_2(e^{j\Omega})=\Omega^2, \forall \Omega \in [-\pi,\pi].\]
\item [(b)] The nonlinear subsystems $\mathbf V_i$ are modeled as third 
order Volterra series, with memory $m=1$, i.e.
\[  ({\mathbf V_j}w)[n]=\sum_{(\alpha(k),\beta(k))}c^j_k
\prod_{l=0}^{1}i[n-l]^{\alpha_l(k)}~
\prod_{l=0}^{1}q[n-l]^{\beta_l(k)},\]
\[\alpha_l(k),\beta_l(k)\in\mathbb Z_+,\qquad
 \sum_{l=0}^1\alpha_l(k)+\sum_{l=0}^1\beta_l(k)\le 3,\]
where $i[n]=\text Re~w[n]$ and $q[n]=\text Im~w[n]$, and 
$(\alpha(k),\beta(k))=(\alpha_0(k),\alpha_1(k),\beta_0(k),\beta_1(k))$.
\end{itemize}
We compare performance of this compensator with the widely used one 
obtained by utilizing simple Volterra series structure \cite{Vuolevi2003}:
\[  ({\mathbf K}w)[n]=\sum_{(\alpha(k),\beta(k))}c_k
\prod_{l=-m_1}^{m_2}i[n-l]^{\alpha_l(k)}~
\prod_{l=-m_1}^{m_2}q[n-l]^{\beta_l(k)},\]
\[\alpha_l(k),\beta_l(k)\in\mathbb Z_+,\qquad
 \sum_{l=-m_1}^{m_2}\alpha_l(k)+\sum_{l=-m_1}^{m_2}\beta_l(k)\le d.\]
Parameters of $\bf K$ which could be varied are forward and backward 
memory depth $m_1$ and $m_2$, respectively, and degree $d$ of this 
model. We consider three cases for different sets of parameter values: 
\begin{itemize}
\item Case 1: $m_1=0,~m_2=2,~d=5$
\item Case 2: $m_1=0,~m_2=4,~d=5$
\item Case 3: $m_1=2,~m_2=2,~d=5$
\end{itemize}

\begin{table}[h]
\caption{Number of coefficients $c_k$ being optimized for different compensator models}
\label{table}
\begin{center}
\begin{tabular}{|c||c|c|}
\hline
Model & \# of $c_k$ & \# of significant $c_k$\\
\hline\hline
New structure & 210 & 141\\
\hline
Volterra 1 & 924 & 177\\
\hline
Volterra 2 & 6006 & 2058\\
\hline
Volterra 3 & 6006 & 1935\\
\hline
\end{tabular}
\end{center}
\end{table}

After fixing the compensator structure, coefficients $c_k$ are obtained by 
applying straightforward least squares optimization. \\
We should emphasize here that fitting has to be done 
for both real and imaginary part of $v[n]$, thus the actual compensator structure 
is twice that depicted in Figure~\ref{fig:Compensator_structure}. \\

\begin{figure}
\center
\includegraphics[scale=.65]{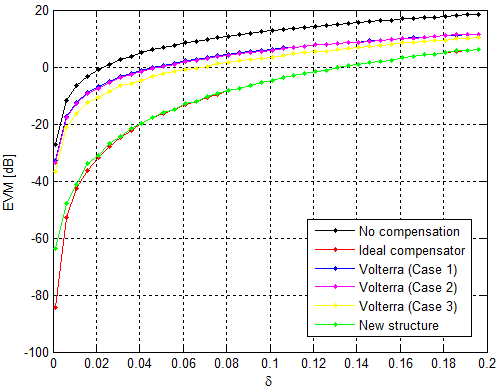}
\caption{Output EVM for different compensator structures}
\label{fig:EVM_comparison}
\end{figure}

Simulation parameters for system {\bf S} are as follows: symbol rate 
$f_{symb}=2\text{MHz}$, carrier frequency $f_c=20\text{MHz}$, 
with 64QAM input symbol sequence. Nonlinear distortion subsystem {\bf F} 
of {\bf S}, used in simulation, is defined in \eqref{eq:distortion}, where the 
delays $\tau_1,\tau_2,\tau_3$ are given by the vector 
$\tau = [0.2T\quad 0.3T \quad 0.4T]$, with $T=1/f_{symb}$. Digital 
simulation of the continuous part of {\bf S} was done by representing 
continuous signals by their discrete counterparts, obtained by sampling 
with high sampling rate $f_s=1000\cdot f_{symb}$. We use a 64QAM 
symbol sequence, with period $N_{symb}=4096$, as an input to {\bf S}. 
This period length is used for generating input/output data for fitting 
coefficients $c_k$, as well as generating input/output data for 
performance validation.  
\\
In Figure~\ref{fig:EVM_comparison} we present EVM obtained for different compensator 
structures, as well as output EVM with no compensation, and case with ideal compensator 
$\mathbf C=\mathbf S^{-1}\approx 2\mathbf I-\mathbf S$. As can be seen from 
Figure~\ref{fig:EVM_comparison}, compensator fitted using the proposed structure 
in Figure~\ref{fig:Compensator_structure}
outperforms other compensators, and gives output EVM almost identical to the ideal
compensator. This result was to be expected, since model in 
Figure~\ref{fig:Compensator_structure} approximates the original system {\bf S} 
very closely, and thus is capable of approximating system $2\mathbf I-\mathbf S$ 
closely as well. This is not the case for compensators modeled with simple Volterra 
series, due to inherently long (or more precisely infinite) memory introduced 
by the LTI part of {\bf S}. Even if we use noncausal Volterra series model 
(i.e. $m_1\not=0$), which 
is expected to capture true dynamics better, we are still unable to get good fitting 
of the system {\bf S}, and consequently of the compensator 
$\mathbf C\approx 2\mathbf I-\mathbf S$.\\
Advantage of the proposed compensator structure is not only in better 
compensation performance, but also in that it achieves better performance 
with much more efficient strucuture. That is, we need far less 
coefficients in order to represent nonlinear part of the compensator, in 
both least squares optimization and actual implementation (Table~\ref{table}).  
In Table~\ref{table} we can see a comparison in the number of coefficients 
between different compensator structures, for nonlinear subsystem parameter 
value $\delta = 0.02$.  Data in the first column is number of coefficients 
(i.e. basis elements) needed for general Volterra model, i.e. coefficients 
which are optimized by least squares. The second column shows actual number 
of coefficients 
used to build compensator. Least squares optimization yields many nonzero 
coefficients, but only subset of those are considered 
significant and thus used in actual compensator implementation. 
Coefficient is considered significant if its value falls above 
a certain treshold $t$, where $t$ is chosen such that increase in EVM after zeroing 
nonsignificant coefficients is not larger than 1\% of the best achievable EVM 
(i.e. when all basis elements are used for building compensator). From 
Table~\ref{table} we can see that for case 3 Volterra structure, 10 times more 
coefficients are needed in order to implement compensator, 
than in the case of our proposed structure. And even when such a large number of 
coefficients is used, its performance is still below the one achieved by this new 
compensator model.

\section{Conclusion}
In this paper, we propose a novel explicit expression of the equivalent baseband model,  
under assumption that the passband nonlinearity can be described by a Volterra series model 
with the fixed degree and memory depth. This result suggests a new, non-obvious, analytically 
motivated structure of digital precompensation of passband nonlinear distortions caused by power 
amplifiers, in digital communication systems. It has been shown that the baseband equivalent 
model is 
a series connection of a fixed degree and short memory 
Volterra model, and a long memory discrete-time LTI system, called reconstruction filter. 
Frequency response of the reconstruction filter 
is shown to be smooth, hence well aproximable by low order polynomials. 
Parameters of such a model (and accordingly of the predistorter) can be obtained by applying 
simple least squares optimization to the input/output data measured from the system, thus 
implying low implementation complexity. State of the art implementations of DPD, have long 
memory requirements in the nonlinear subsystem, but structure of our baseband equivalent 
model suggests that the long memory requirements can be shifted from the nonlinear part 
to the LTI part, which consists of FIR filters and is easy to implement in 
digital circuits, giving it advantage of much lower complexity. We also argued that this 
baseband model, and its corresponding DPD structure, can be readily extended to OFDM 
modulation. 
Simulation results have shown that by using this new DPD structure, significant reduction in 
nonlinear distortion caused by the RF PA can be achieved, while utilizing full frequency 
band, and thus effectively using maximal input symbol rate. 

\section*{Acknowledgment}
The authors are grateful to Dr. Yehuda Avniel for bringing researchers from
vastly different backgrounds to work together on the tasks that led to the writing of
this paper.


\end{document}